\documentclass[a4paper,10pt]{iopart}
\usepackage[utf8x]{inputenc}
\usepackage{color}
\usepackage{graphicx}
\usepackage{amssymb}
\usepackage[bookmarks, bookmarksopen, bookmarksnumbered]{hyperref}

\newcommand{\dd}{\ensuremath{\textrm{d}}}

\newcommand{\RealNum}{\ensuremath{{\bf R}}}

\newcommand{\Nn}{\ensuremath{{\cal N }}}
\newcommand{\beq}{\begin{equation}}
\newcommand{\eeq}{\end{equation}}
\newcommand{\bea}{\begin{eqnarray}}
\newcommand{\eea}{\end{eqnarray}}
\newcommand{\bit}{\begin{itemize}}
\newcommand{\eit}{\end{itemize}}
\newcommand{\bfi}{\begin{figure}}
\newcommand{\efi}{\end{figure}}
\newcommand{\bfic}{\begin{figure*}}
\newcommand{\efic}{\end{figure*}}
\newcommand{\bce}{\begin{center}}
\newcommand{\ece}{\end{center}}
\newcommand{\bt}{\begin{table}}
\newcommand{\et}{\end{table}}
\newcommand{\btb}{\begin{tabular}}
\newcommand{\etb}{\end{tabular}}

\newcommand{\eff}{\ensuremath{{\rm eff}}}

\newcommand{\qed}{\nobreak \ifvmode \relax \else
      \ifdim\lastskip<1.5em \hskip-\lastskip
      \hskip1.5em plus0em minus0.5em \fi \nobreak
      \vrule height0.75em width0.5em depth0.25em\fi}

\begin{document}

\hspace{4.5in} \mbox{UW-ThPh-2012-17}

\hspace{4.5in} \mbox{AEI-2012-36}

\title{Evolution of a periodic eight-black-hole lattice in numerical relativity}

\author{Eloisa Bentivegna}
\address{
Max-Planck-Institut f\"ur Gravitationsphysik\\
Albert-Einstein-Institut \\ 
Am M\"uhlenberg 1, D-14476 Golm \\
Germany}

\author{Miko\l{}aj Korzy\'nski}
\address{
Gravitational Physics \\
Faculty of Physics \\
University of Vienna, A-1090 Vienna \\
Austria}

\begin{abstract}
The idea of black-hole lattices as models for the large-scale structure of
the universe has been under scrutiny for several decades, and some of the
properties of these systems have been elucidated recently in the context of
the problem of cosmological backreaction. The complete, three-dimensional
and fully relativistic evolution of these system has, however, never been tackled. 
We explicitly construct the first of these solutions by numerically integrating
Einstein's equation in the case of an eight-black-hole lattice with the topology 
of $S^3$.
\end{abstract}


\section{Introduction}
The vast amount of large-scale cosmological data collected in recent
decades has shaped a generally coherent picture of our universe~\cite{CervantesCota:2011pn},
where the thermodynamics and nucleosynthesis in the early universe, 
the generation of the seeds of cosmic structure and their subsequent
evolution all fit together in a simple framework based on remarkably 
few principles. This success, however, comes at the price of accepting
the existence of a dark sector, i.e.~two fluids, dark 
matter~\cite{sanders2010dark} and dark energy~\cite{wang2010dark,amendola2010dark}, 
which have rather peculiar physical properties, have an uncertain
collocation within the Standard Model, and have never been observed 
in terrestrial laboratories despite accounting for over 95\% of the energy
density of the universe.

Whilst this result could very well point to the existence of new physical 
constituents or principles, 
the possibility that 
the current way we model the inhomogeneous universe be too elementary
(a possibility that was advanced for the first time in~\cite{Ellis:1987zz})
has now resurfaced~\cite{0264-9381-28-16-160301}. In particular, the question of the extent to which cosmic
inhomogeneities can dress the value of the cosmological parameters
is under scrutiny in a variety of approaches~\cite{Korzynski:2009db,
0264-9381-28-16-164002,0264-9381-28-16-164007,0264-9381-28-16-164011, Brannlund:2010rs,Wiltshire:2009ys,Wiltshire:2007zr} 
(see also the review articles \cite{Ellis:2011, Wiltshire:2011zp} and references therein).

An interesting class of models that has been studied for some time is 
that of regular lattices of black holes~\cite{RevModPhys.29.432,Clifton:2009jw}
\footnote{Notice that the 
usual definition of a black hole in an asymptotically-flat spacetime
is inapplicable to these spaces. Here, by black hole we denote a 
spacetime region surrounded by a marginally outer-trapped tube (MOTT).}. These 
representations of
our universe avoid the issues related to the behavior of relativistic
fluids (and, in particular, the corresponding singularities);
one could argue that they also represent a more realistic picture
of our universe, composed of a collection of pointlike objects surrounded
by vacuum rather than a homogeneous and isotropic fluid with small-scale
perturbations.

Collections of black holes also have the added benefit to be one of
numerical relativity's routine application areas~\cite{Pfeiffer:2012pc}, 
from which formalisms, tools and practices can be readily imported.
In this work, we construct the initial data and simulate the evolution
of a special sort of black hole lattices, those with extrinsic curvature  
that is initially zero. In section~\ref{sec:construction}, we illustrate
how to construct exact initial data for a generic black-hole lattice based on the 
usual Lichnerowicz-York framework~\cite{lichnerowicz:1944,York:1971hw}. 
We then discuss a coordinate 
transformation that simplifies the numerical treatment in section~\ref{sec:stereo},
illustrate the details of the evolution in section~\ref{sec:evolution},
interpret the results and contrast them to the homogeneous and 
isotropic class in~\ref{sec:contrast}, and finally conclude in section~\ref{sec:conclusions}.
Unless otherwise stated, greek indices run from 0 to 4, latin indices run from 1 to 3, and we set $G=c=1$.

\section{Constructing a periodic black-hole lattice}
\label{sec:construction}
Given the standard 3+1 split of the metric tensor into the
spatial metric $\gamma_{ij}$ and extrinsic curvature $K_{ij}$,
initial data for the gravitational field can be generated by
solving the Hamiltonian and momentum constraints, which
read:
\bea
\label{eq:constraints}
R+K^2-K_{ij}K^{ij} = 16\pi G \rho \\
D_j K^j_i-D_j K    = 8 \pi G j_i
\eea
where $R$ is the scalar curvature of the spatial metric and 
$D_i$ represents the covariant derivative associated with
$\gamma_{ij}$; $\rho=n^\mu n^\nu T_{\mu\nu}$ and $j^i=-\gamma^{ij} n^\mu T_{j\mu}$ represent 
the energy and momentum density respectively.

Let us assume that $j_i$ vanishes.
A powerful scheme to generate solutions of this system is
the conformal transverse-traceless framework, which entails the
introduction of a conformal transformation in the spatial
metric, along with the separation of the extrinsic curvature
into its trace $K$ and traceless part $A_{ij}$:
\bea
\label{eq:YorkID}
 \gamma_{ij} &=& \psi^4\,\tilde \gamma_{ij} \\
 K_{ij} &=& \frac{K}{3}\,\gamma_{ij} + A_{ij}
\eea
In terms of these, the constraints take the form:
\begin{eqnarray}
\label{eq:CTTconstraints}
 \tilde \Delta \psi - \frac{\tilde R}{8}\,\psi - \frac{K^2}{12}\,\psi^5 + \frac{1}{8} {\tilde A}_{ij} {\tilde A}^{ij} \psi^{-7} = -2\pi G\,\psi^5\,\rho \\
 \tilde D_i \tilde A^{ij} - \frac{2}{3} \psi^6 \tilde \gamma^{ij} \tilde D_i K = 0
\end{eqnarray}
$\tilde \Delta$ being the laplacian operator of the conformal metric 
$\tilde \gamma_{ij}$, and $\tilde A_{ij}$ being related to $A_{ij}$ by 
$\tilde A_{ij}=\psi^2 A_{ij}$~\footnote{A different scaling of $\rho$ is preferable if we want to solve the initial value problem for some types of matter, but
this plays no role in our argument.}.

Let us focus on the Hamiltonian constraint.
We would like to solve this equation with periodic boundary conditions. We also allow for matter content in form of  
ordinary matter $\rho$ as well as in ``punctures'', i.e.~singularities in $\psi$ of the form $m_i/2r$, $m_i > 0$. 
It can be easily proven that, unlike in the asymptotically-flat case, if we set both $K_{ij}$ and $R$ to zero, then this is a slice of Minkowski spacetime. 

To see this, let us first integrate both sides of equation (\ref{eq:CTTconstraints}) over the fundamental
cell $D$ (which, for the sake of illustration, we will assume cubical) of the desired lattice, 
with small balls around the punctures excised at the surfaces $S_i$ (see Figure~\ref{fig:cell}).
\bfi
\bce
\includegraphics[width=1.0\textwidth, trim=20 100 0 0, clip=true]{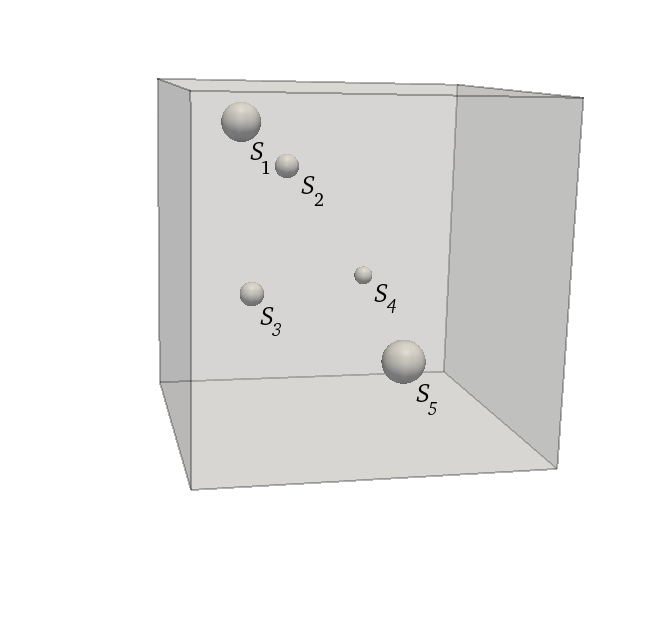}
\caption{The elementary cell of a periodic 3-space containing a number of punctures and
corresponding inner boundaries, e.g.~in the shape of spherical surfaces $S_i$.\label{fig:cell}}
\ece
\efi
The volume integral of $\tilde \Delta \psi$ can then be turned into a surface integral on
the $S_i$ alone, as $\tilde \Delta \psi$ identically vanishes on the periodic boundary.
We thus obtain:
\begin{eqnarray}
\label{eq:IntHamiltonianconstaint}
\nonumber
\int_D &&\left(\frac{\tilde R}{8}\,\psi + \frac{1}{12}\,K^2\,\psi^5 - \frac{1}{8}\psi^{-7}\,\tilde A^{ij}\,\tilde A_{ij} 
\right) \sqrt{\tilde\gamma}\,d^3 x = \\
= &&2\pi G\,\left(\int_D \rho\psi^5\,\sqrt{\tilde\gamma}\,d^3 x + \Sigma_{i=1}^{N} m_i\right).
\end{eqnarray}
On the right hand side, which is manifestly positive, we have the total energy content of the cell both in form of a continuous distribution as well as
in the punctures. If we set both $\tilde R$ and $K$ to zero, the equation becomes impossible to satisfy unless the matter content vanishes as well.
Thus, for non-zero $m_i$, we need to admit either a non-zero extrinsic curvature $K$ or a positive
spatial scalar curvature $\tilde R$. 

In this work we concentrate
on the second case, setting $K$ to vanish. In this case the momentum constraint is trivally satisfied and the 
Hamiltonian constraint remains linear in $\psi$, which allows for constructing multiple-black-hole
solutions by superposition. 

We would like the conformal metric $\tilde\gamma$ to be periodic just like the physical one. The simplest way to 
ensure that it admits a discrete symmetry is to assume that it is a hyperbolic, spherical or flat metric. 
Since the integral of $\tilde R$ must be positive, $\tilde\gamma$ must be the metric of a round 3-sphere. This condition limits both the form of
the metric tensor and the topology of the lattice~\footnote{In a more general setting, the scalar curvature does not have a definite sign and the integral condition (\ref{eq:IntHamiltonianconstaint}) gives little information. Nevertheless, thanks to the Yamabe theorem, $\tilde \gamma$ must be conformally equivalent to a constant-curvature metric. If we rewrite
equation (\ref{eq:IntHamiltonianconstaint}) in terms of this metric, it obviously becomes a condition for the sign of the Yamabe energy
 $\mathcal{E}(\tilde\gamma)$. }.

Let us note that a similar reasoning applies to the Friedmann-Lema\^itre-Robertson-Walker (FLRW) models: the Hamiltonian constraint in this class reads
\bea
\frac{R}{8} + \frac{K^2}{12} = 2\pi G \rho.
\eea
Since the RHS, representing the matter content, is manifestly positive, then either we must have non-vanishing Hubble parameter $K$, or a positive
curvature $R$ (or both). Thus, if we assume $K=0$, then there is a similarly strong restriction on the metric tensor and the topology of the constant-time slices.

\subsection{Punctures on $S^3$}
\label{sec:S3punct}

Following~\cite{springerlink:10.1007/BF01889418}~\footnote{The analysis of this initial-data construction in 
the context of the backreaction problem has also recently appeared in~\cite{Clifton:2012qh}},
we consider puncture-like solutions of the Hamiltonian constraint
when $\tilde \gamma_{ij}$ and $\tilde R$ are respectively the metric tensor $\gamma^S_{ij}$ and the scalar curvature of the
round 3-sphere: 
\begin{eqnarray}
\label{eq:hamS3}
 \tilde \Delta\,\psi - \frac{\tilde R}{8}\,\psi = 0
\end{eqnarray}
We fix coordinates on $S^3$ such that:
\beq
\gamma^S = \dd\lambda^2 + \sin^2\lambda\,\left(\dd\theta^2 + \sin^2\theta\,\dd\varphi^2\right)
\eeq
Let us imagine that the sphere is embedded in $\RealNum^4$ with the equation
$\left(X^1\right)^2 + \left(X^2\right)^2 + \left(X^3\right)^2 + \left(X^4\right)^2 = 1$; a bar over a capital letter denotes 
a vector in this space.

Equation (\ref{eq:hamS3}) has no regular solutions, but it is straightforward to find its solutions with a puncture-type singularity:
\begin{eqnarray}
 \psi(\lambda) = \frac{A}{\sin \lambda/2}.
\end{eqnarray}
or in Cartesian coordinates
\begin{eqnarray}
 \psi(\bar X) = A\sqrt{\frac{2}{1 - \bar X\cdot \bar N}}, \\
 \bar N = (0,0,0,1)
\end{eqnarray}

We can easily superimpose $N$ such punctures
centered at chosen locations $\bar N_i \in S^3$:
\begin{eqnarray}
 \psi(\bar X) = \sum_{i=1}^{N} \frac{A_i}{\sin \lambda_i/2} = \sum_i A_i\,\sqrt{\frac{2}{1 - \bar X\cdot \bar N_i}} \label{eq:phiii}
\end{eqnarray}
The parameters $A_i > 0$ measure the singular part 
of the solution at the points $\bar N_i$: the leading part behaves like
$2A_i / \lambda_i$.

Notice that, if one seeks only the \emph{regular} arrangements of black holes on $S^3$, 
there are only six possible values of $N$, corresponding to the six regular tessellations
of the 3-sphere: $N=5,8,16,24,120,600$.
In the following, we concentrate on the 8-vertex, 16-cell solution, where the puncture
locations are given by:
\begin{eqnarray}
\label{eq:8vertex}
 \bar N_1 &=& \left(1, 0, 0, 0\right), \nonumber \\
 \bar N_2 &=& \left(-1,0,0,0\right), \nonumber\\
 \bar N_3 &=& (0,1,0,0) , \nonumber\\
 \bar N_4 &=& (0,-1,0,0) , \nonumber\\
 \bar N_5 &=& \left(0, 0, 1, 0\right),\nonumber\\
 \bar N_6 &=& \left( 0,0,-1,0\right),\nonumber\\
 \bar N_7 &=& (0,0,0,1) , \nonumber\\
 \bar N_8 &=& (0,0,0,-1).
\end{eqnarray}
and all $A_i = 1$.
The configuration obviously has the symmetry of a 16-cell. In particular,
it has a discrete group of symmetries generated by $\pi/2$ rotations around all pairs of axes of $\RealNum^4$ and reflections about all four hyperplanes perpendicular
to the axes. 
The elementary cell in this pattern is cubical in shape, i.e. it has 6 faces, 8 edges and 8 vertices, at which exactly 4 edges meet. 
All edges lie on great circles of $S^3$ and their length
is equal to 168.343.

\subsection{Stereographic projection of $S^3$}
\label{sec:stereo}
Since it is easier to perform the evolution of asymptotically-flat data as opposed to data on a sphere, we 
employ the stereographic projection from the top of the sphere into $\RealNum^3$, given by 
\begin{eqnarray}
\label{eq:stereo}
 x^i = \frac{2X^i}{1-X^4}
\end{eqnarray}
It is well known that the projection is a conformal mapping in the sense that
\begin{eqnarray}
 \gamma^S_{ij} = \left(|\vec x|^2/4 + 1\right)^{-2}\,\delta_{ij}
\end{eqnarray}
or
\begin{eqnarray}
 \gamma^S_{ij} = \left(\sin\frac{\lambda}{2}\right)^{-4}\,\delta_{ij} = \left(\frac{2}{1 - \cos\lambda}\right)^2\,\delta_{ij}
\end{eqnarray}
where $\delta_{ij}$ is the flat metric. The physical metric (\ref{eq:phiii}) projected down to $\RealNum^3$ takes the form of
\begin{eqnarray}
 \psi^4\,\gamma^S_{ij} &=& \left(A_1\right)^4\,\tilde\psi^4 \,\delta_{ij} \\
 \tilde\psi(\vec x) &=& 1 + \sum_{i=2}^{N}\,\frac{2A_i\,\sqrt{1+|\vec n_i|^2/4}}{A_1}\cdot\frac{1}{\left|\vec x - \vec n_i\right|}
\end{eqnarray}
Thus the potential consist of $N-1$ punctures of $1/r$ type, one of the punctures having been projected out to infinity. Note that the physical metric involves the (scale-setting) 
factor $(A_1)^4$. We can absorb it by introducing new, rescaled coordinates $\vec y = (A_1)^2\,\vec x$. The projected conformal
factor takes now the form of
\begin{eqnarray}
\label{eq:confac}
 \tilde\psi(\vec y) &=& 1 + \sum_{i=2}^{N}\,\frac{2A_i\,A_1\,\sqrt{1+\frac{|\vec {\cal N}_i|^2}{4}}}{\left|\vec y - \vec {\cal N}_i\right|} = \\
 &=&1 + \sum_{i=2}^{N} \frac{m_i}{2\left|\vec y - \vec {\cal N}_i\right|} \label{eqtildephi2}
\end{eqnarray}
with rescaled positions of the punctures $\vec {\cal N}_i = (A_1)^2\,\vec n_i$. 
The mass parameters of the punctures take the form of
\begin{eqnarray}
 m_i = 4A_i\,A_1\,\sqrt{1+\frac{|\vec {\cal N}_i|^2}{4}}.
\end{eqnarray}
They have the dimension of mass, but do not correspond exactly to the ADM mass of the individual punctures measured at their infinities. 
For the black hole at $\bar {\cal N}_1$, for instance, the ADM mass is equal to
\begin{eqnarray}
 M^{\rm{ADM}}_1 = \sum_{i=2}^N \,m_i = 4A_1\,\sum_{i=2}^N\,A_i\,\sqrt{1+\frac{|\vec {\cal N}_i|^2}{4}}.
\end{eqnarray}
This can also be expressed in terms of the original solution $\psi$:
\begin{eqnarray}
 M^{\rm{ADM}}_1 = 4A_1\,\sum_{i=1}^N \, A_i\sqrt{\frac{2}{1 - \bar N_i\cdot \bar N_1}}.
\end{eqnarray}
Analogous relations hold for other black holes.

If we project the 8-vertex solution (\ref{eq:8vertex}) down to $\RealNum^3$, it becomes an
asymptotically-flat configuration described by (\ref{eqtildephi2}) with 7 punctures at points 
\begin{eqnarray}
 \vec\Nn_2 &=& \left(0,0,0\right),\\
 \vec\Nn_3 &=& (2,0,0) , \\
 \vec\Nn_4 &=& (-2,0,0) , \\
 \vec\Nn_5 &=& \left( 0, 2, 0\right),\\
 \vec\Nn_6 &=& \left( 0,-2,0\right),\\
 \vec\Nn_7 &=& (0,0,2) , \\
 \vec\Nn_8 &=& (0,0,-2).
\end{eqnarray}
and with mass parameters $m_2 = 4$ and $m_3,\dots,m_7 = 4\sqrt{2}$. 
The vertices, edges and marginally outer-trapped surfaces (MOTSs) projected to $\RealNum^3$ are presented on Figure~\ref{fig:proj}. 
\bfi
\bce
\includegraphics[width=0.7\textwidth]{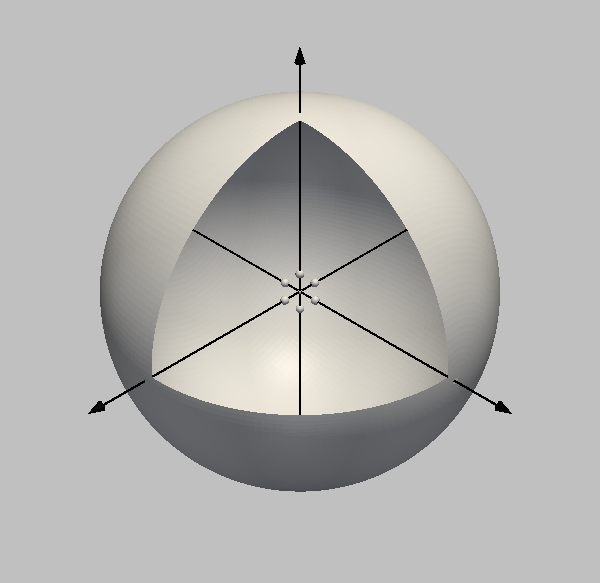}

\vspace{0.5in}

\includegraphics[width=0.7\textwidth]{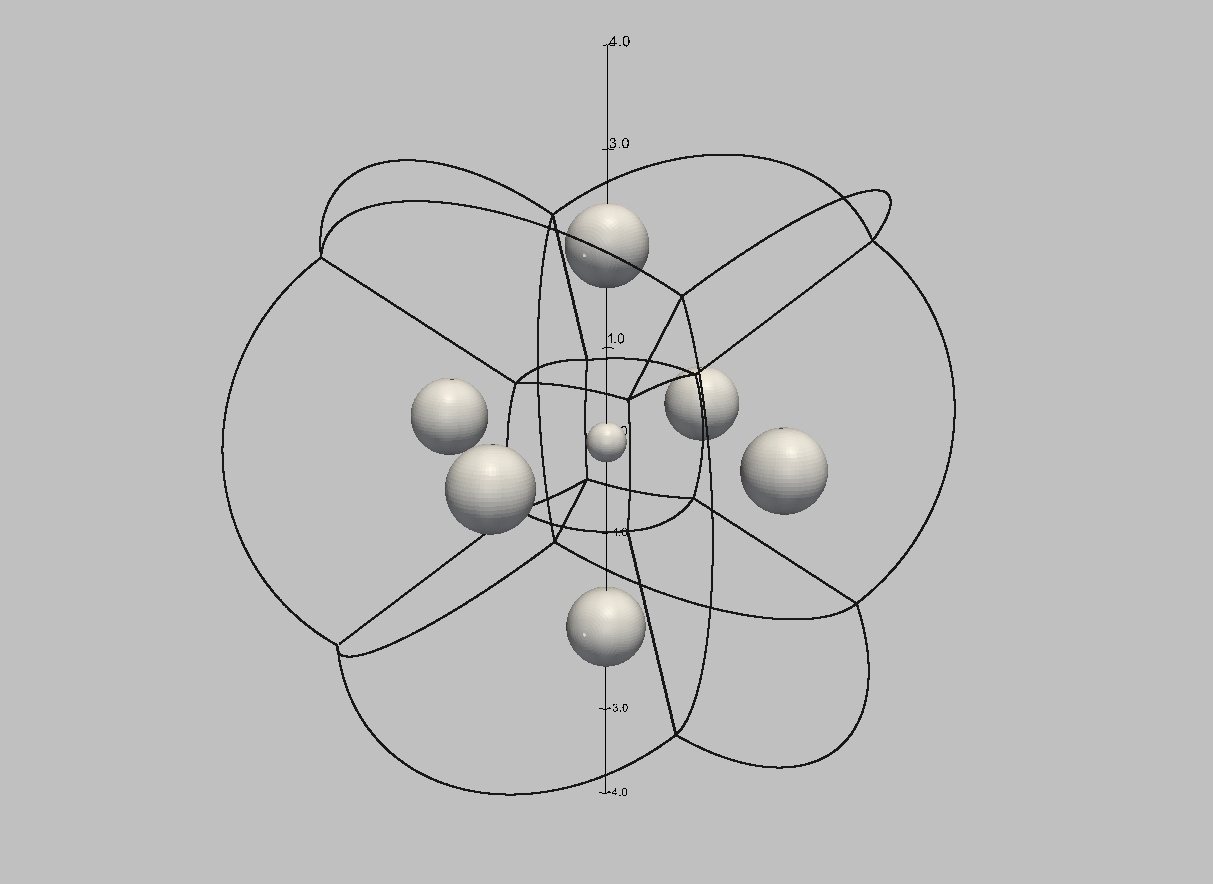}
\ece
\caption{Initial MOTSs and elementary cells for the 8-black-hole configuration,
projected to $\RealNum^3$. The
marginal surface corresponding to the black hole at infinity encompasses the whole configuration. 
Note that the 8 cubical lattice cells are isometric after the conformal rescaling.\label{fig:proj}}
\efi

\section{Evolution} 
\label{sec:evolution}

We perform the three-dimensional evolution of this initial-data set using the
Einstein Toolkit~\cite{Loffler:2011ay}; in particular, we use the \texttt{McLachlan} 
code~\cite{mclachlan,kranc} to perform a finite-difference evolution of the Einstein's
equation with adaptive-mesh-refinement capabilities
provided by \texttt{Carpet}~\cite{carpet}.
We have also made use of \texttt{AHFinderDirect}~\cite{Thornburg:2003sf} 
to search for trapped surfaces.

Details of the evolution scheme can be found in~\ref{app:bssn}; the
numerical error analysis is performed in~\ref{app:res}.
We evolve the initial data presented in section~\ref{sec:stereo} 
in a cubic box of side $40M$ (here and in the following, we will adopt $M=m_2/4$ 
as the unit of mass, length and time), with a spacing $\Delta_0=1M$. We refine
the grid at the seven black-hole locations using eight concentric grids for
each location, down to a resolution of $\Delta_8=\Delta_0/2^8=0.00390625M$.

The eight initial MOTSs have been shown in Figure~\ref{fig:proj}
\footnote{We really only track the surfaces corresponding to the black holes
at the origin, at infinity, and on the positive $x$-axis. The locations and shapes of the
remaining five surfaces, included in the plots for clarity, are obtained
by symmetry arguments.}.
Notice that, due to the stereographic projection, the large marginal surface
initially at a radius of $20M$ is {\it inner} trapped, rather than outer trapped, 
i.e.~it is the expansion of the {\it ingoing} null normal that vanishes
while the expansion of the outer normal is positive. It is therefore
\emph{not} a common apparent horizon of the sort usually encountered in
binary mergers.
This has an interesting side effect: the outer boundary conditions are causally
disconnected from the region enclosed by this surface.

We find that the MOTSs at $(\pm 2, 0, 0)$, $(0, \pm 2, 0)$ and $(0, 0, \pm 2)$ take 
approximately $130M$ (in coordinate time) to merge to the MOTS at the origin, and approximately $170M$ to merge 
to the larger, inner-trapped one, as illustrated in Figure~\ref{fig:merger} (the asymmetry here
is due to the non-uniform numerical slicing). The evolution of the 
$z=0$ sections of the marginal surfaces are shown in Figure~\ref{fig:horizons}, while the
mean coordinate radii and masses for the the black holes initially at (0,0,0), (2,2,2) and 
infinity are plotted in 
Figures~\ref{fig:mass} and \ref{fig:mradius}.
\bfi
\bce
\begingroup
  \makeatletter
  \providecommand\color[2][]{%
    \GenericError{(gnuplot) \space\space\space\@spaces}{%
      Package color not loaded in conjunction with
      terminal option `colourtext'%
    }{See the gnuplot documentation for explanation.%
    }{Either use 'blacktext' in gnuplot or load the package
      color.sty in LaTeX.}%
    \renewcommand\color[2][]{}%
  }%
  \providecommand\includegraphics[2][]{%
    \GenericError{(gnuplot) \space\space\space\@spaces}{%
      Package graphicx or graphics not loaded%
    }{See the gnuplot documentation for explanation.%
    }{The gnuplot epslatex terminal needs graphicx.sty or graphics.sty.}%
    \renewcommand\includegraphics[2][]{}%
  }%
  \providecommand\rotatebox[2]{#2}%
  \@ifundefined{ifGPcolor}{%
    \newif\ifGPcolor
    \GPcolorfalse
  }{}%
  \@ifundefined{ifGPblacktext}{%
    \newif\ifGPblacktext
    \GPblacktexttrue
  }{}%
  \let\gplgaddtomacro\g@addto@macro
  \gdef\gplbacktext{}%
  \gdef\gplfronttext{}%
  \makeatother
  \ifGPblacktext
    \def\colorrgb#1{}%
    \def\colorgray#1{}%
  \else
    \ifGPcolor
      \def\colorrgb#1{\color[rgb]{#1}}%
      \def\colorgray#1{\color[gray]{#1}}%
      \expandafter\def\csname LTw\endcsname{\color{white}}%
      \expandafter\def\csname LTb\endcsname{\color{black}}%
      \expandafter\def\csname LTa\endcsname{\color{black}}%
      \expandafter\def\csname LT0\endcsname{\color[rgb]{1,0,0}}%
      \expandafter\def\csname LT1\endcsname{\color[rgb]{0,1,0}}%
      \expandafter\def\csname LT2\endcsname{\color[rgb]{0,0,1}}%
      \expandafter\def\csname LT3\endcsname{\color[rgb]{1,0,1}}%
      \expandafter\def\csname LT4\endcsname{\color[rgb]{0,1,1}}%
      \expandafter\def\csname LT5\endcsname{\color[rgb]{1,1,0}}%
      \expandafter\def\csname LT6\endcsname{\color[rgb]{0,0,0}}%
      \expandafter\def\csname LT7\endcsname{\color[rgb]{1,0.3,0}}%
      \expandafter\def\csname LT8\endcsname{\color[rgb]{0.5,0.5,0.5}}%
    \else
      \def\colorrgb#1{\color{black}}%
      \def\colorgray#1{\color[gray]{#1}}%
      \expandafter\def\csname LTw\endcsname{\color{white}}%
      \expandafter\def\csname LTb\endcsname{\color{black}}%
      \expandafter\def\csname LTa\endcsname{\color{black}}%
      \expandafter\def\csname LT0\endcsname{\color{black}}%
      \expandafter\def\csname LT1\endcsname{\color{black}}%
      \expandafter\def\csname LT2\endcsname{\color{black}}%
      \expandafter\def\csname LT3\endcsname{\color{black}}%
      \expandafter\def\csname LT4\endcsname{\color{black}}%
      \expandafter\def\csname LT5\endcsname{\color{black}}%
      \expandafter\def\csname LT6\endcsname{\color{black}}%
      \expandafter\def\csname LT7\endcsname{\color{black}}%
      \expandafter\def\csname LT8\endcsname{\color{black}}%
    \fi
  \fi
  \setlength{\unitlength}{0.0500bp}%
  \begin{picture}(7200.00,5040.00)%
    \gplgaddtomacro\gplbacktext{%
      \csname LTb\endcsname%
      \put(1707,704){\makebox(0,0)[r]{\strut{}-20}}%
      \put(1707,1213){\makebox(0,0)[r]{\strut{}-15}}%
      \put(1707,1722){\makebox(0,0)[r]{\strut{}-10}}%
      \put(1707,2231){\makebox(0,0)[r]{\strut{}-5}}%
      \put(1707,2740){\makebox(0,0)[r]{\strut{} 0}}%
      \put(1707,3248){\makebox(0,0)[r]{\strut{} 5}}%
      \put(1707,3757){\makebox(0,0)[r]{\strut{} 10}}%
      \put(1707,4266){\makebox(0,0)[r]{\strut{} 15}}%
      \put(1707,4775){\makebox(0,0)[r]{\strut{} 20}}%
      \put(1839,484){\makebox(0,0){\strut{}-20}}%
      \put(2348,484){\makebox(0,0){\strut{}-15}}%
      \put(2857,484){\makebox(0,0){\strut{}-10}}%
      \put(3366,484){\makebox(0,0){\strut{}-5}}%
      \put(3875,484){\makebox(0,0){\strut{} 0}}%
      \put(4383,484){\makebox(0,0){\strut{} 5}}%
      \put(4892,484){\makebox(0,0){\strut{} 10}}%
      \put(5401,484){\makebox(0,0){\strut{} 15}}%
      \put(5910,484){\makebox(0,0){\strut{} 20}}%
      \put(1069,2739){\rotatebox{-270}{\makebox(0,0){\strut{}$y/M$}}}%
      \put(3874,154){\makebox(0,0){\strut{}$x/M$}}%
    }%
    \gplgaddtomacro\gplfronttext{%
      \csname LTb\endcsname%
      \put(4923,4602){\makebox(0,0)[r]{\strut{}t=0M}}%
      \csname LTb\endcsname%
      \put(4923,4382){\makebox(0,0)[r]{\strut{}t=25M}}%
      \csname LTb\endcsname%
      \put(4923,4162){\makebox(0,0)[r]{\strut{}t=50M}}%
      \csname LTb\endcsname%
      \put(4923,3942){\makebox(0,0)[r]{\strut{}t=75M}}%
      \csname LTb\endcsname%
      \put(4923,3722){\makebox(0,0)[r]{\strut{}t=100M}}%
      \csname LTb\endcsname%
      \put(4923,3502){\makebox(0,0)[r]{\strut{}t=125M}}%
      \csname LTb\endcsname%
      \put(4923,3282){\makebox(0,0)[r]{\strut{}t=150M}}%
    }%
    \gplbacktext
    \put(0,0){\includegraphics{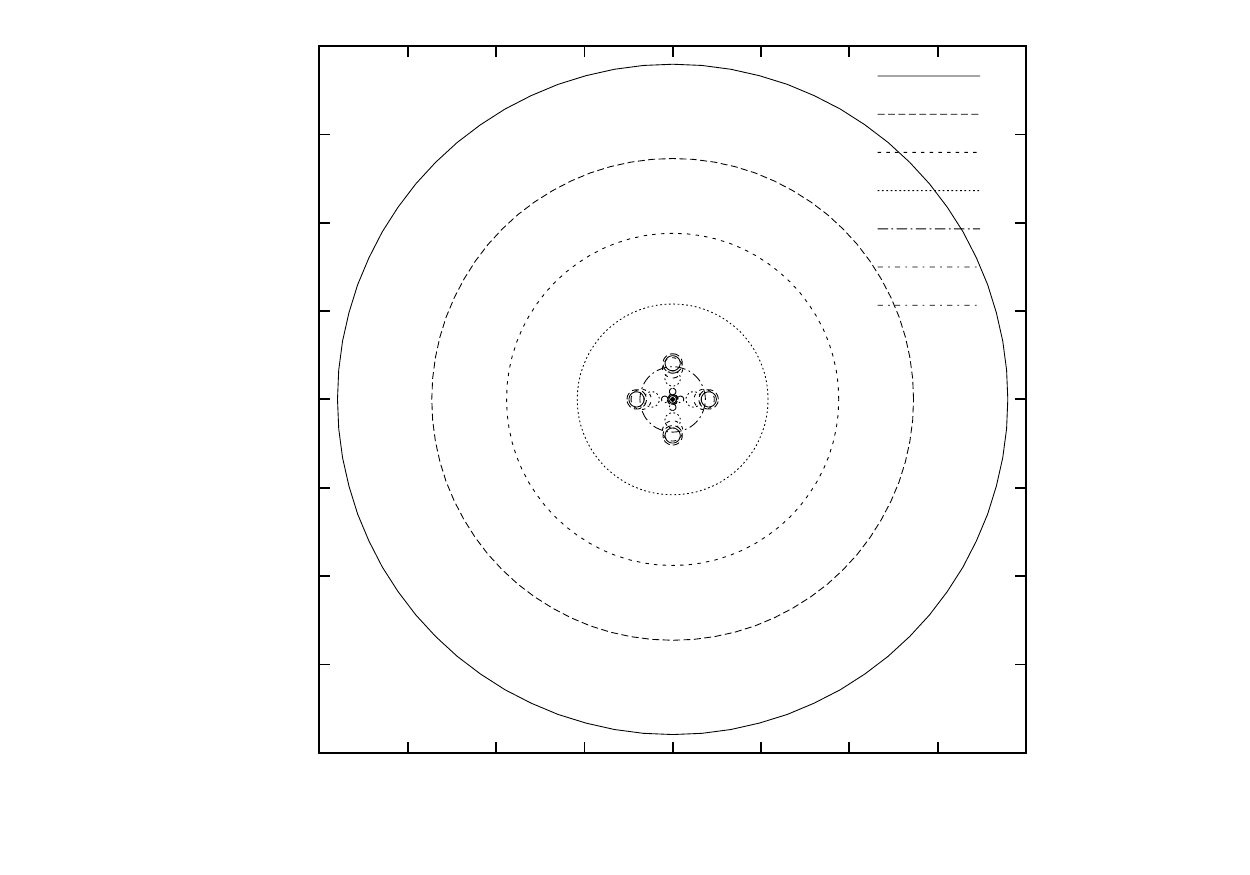}}%
    \gplfronttext
  \end{picture}%
\endgroup

\begingroup
  \makeatletter
  \providecommand\color[2][]{%
    \GenericError{(gnuplot) \space\space\space\@spaces}{%
      Package color not loaded in conjunction with
      terminal option `colourtext'%
    }{See the gnuplot documentation for explanation.%
    }{Either use 'blacktext' in gnuplot or load the package
      color.sty in LaTeX.}%
    \renewcommand\color[2][]{}%
  }%
  \providecommand\includegraphics[2][]{%
    \GenericError{(gnuplot) \space\space\space\@spaces}{%
      Package graphicx or graphics not loaded%
    }{See the gnuplot documentation for explanation.%
    }{The gnuplot epslatex terminal needs graphicx.sty or graphics.sty.}%
    \renewcommand\includegraphics[2][]{}%
  }%
  \providecommand\rotatebox[2]{#2}%
  \@ifundefined{ifGPcolor}{%
    \newif\ifGPcolor
    \GPcolorfalse
  }{}%
  \@ifundefined{ifGPblacktext}{%
    \newif\ifGPblacktext
    \GPblacktexttrue
  }{}%
  \let\gplgaddtomacro\g@addto@macro
  \gdef\gplbacktext{}%
  \gdef\gplfronttext{}%
  \makeatother
  \ifGPblacktext
    \def\colorrgb#1{}%
    \def\colorgray#1{}%
  \else
    \ifGPcolor
      \def\colorrgb#1{\color[rgb]{#1}}%
      \def\colorgray#1{\color[gray]{#1}}%
      \expandafter\def\csname LTw\endcsname{\color{white}}%
      \expandafter\def\csname LTb\endcsname{\color{black}}%
      \expandafter\def\csname LTa\endcsname{\color{black}}%
      \expandafter\def\csname LT0\endcsname{\color[rgb]{1,0,0}}%
      \expandafter\def\csname LT1\endcsname{\color[rgb]{0,1,0}}%
      \expandafter\def\csname LT2\endcsname{\color[rgb]{0,0,1}}%
      \expandafter\def\csname LT3\endcsname{\color[rgb]{1,0,1}}%
      \expandafter\def\csname LT4\endcsname{\color[rgb]{0,1,1}}%
      \expandafter\def\csname LT5\endcsname{\color[rgb]{1,1,0}}%
      \expandafter\def\csname LT6\endcsname{\color[rgb]{0,0,0}}%
      \expandafter\def\csname LT7\endcsname{\color[rgb]{1,0.3,0}}%
      \expandafter\def\csname LT8\endcsname{\color[rgb]{0.5,0.5,0.5}}%
    \else
      \def\colorrgb#1{\color{black}}%
      \def\colorgray#1{\color[gray]{#1}}%
      \expandafter\def\csname LTw\endcsname{\color{white}}%
      \expandafter\def\csname LTb\endcsname{\color{black}}%
      \expandafter\def\csname LTa\endcsname{\color{black}}%
      \expandafter\def\csname LT0\endcsname{\color{black}}%
      \expandafter\def\csname LT1\endcsname{\color{black}}%
      \expandafter\def\csname LT2\endcsname{\color{black}}%
      \expandafter\def\csname LT3\endcsname{\color{black}}%
      \expandafter\def\csname LT4\endcsname{\color{black}}%
      \expandafter\def\csname LT5\endcsname{\color{black}}%
      \expandafter\def\csname LT6\endcsname{\color{black}}%
      \expandafter\def\csname LT7\endcsname{\color{black}}%
      \expandafter\def\csname LT8\endcsname{\color{black}}%
    \fi
  \fi
  \setlength{\unitlength}{0.0500bp}%
  \begin{picture}(7200.00,5040.00)%
    \gplgaddtomacro\gplbacktext{%
      \csname LTb\endcsname%
      \put(1641,704){\makebox(0,0)[r]{\strut{}-3}}%
      \put(1641,1383){\makebox(0,0)[r]{\strut{}-2}}%
      \put(1641,2061){\makebox(0,0)[r]{\strut{}-1}}%
      \put(1641,2740){\makebox(0,0)[r]{\strut{} 0}}%
      \put(1641,3418){\makebox(0,0)[r]{\strut{} 1}}%
      \put(1641,4097){\makebox(0,0)[r]{\strut{} 2}}%
      \put(1641,4775){\makebox(0,0)[r]{\strut{} 3}}%
      \put(1773,484){\makebox(0,0){\strut{}-3}}%
      \put(2452,484){\makebox(0,0){\strut{}-2}}%
      \put(3130,484){\makebox(0,0){\strut{}-1}}%
      \put(3809,484){\makebox(0,0){\strut{} 0}}%
      \put(4487,484){\makebox(0,0){\strut{} 1}}%
      \put(5166,484){\makebox(0,0){\strut{} 2}}%
      \put(5844,484){\makebox(0,0){\strut{} 3}}%
      \put(1135,2739){\rotatebox{-270}{\makebox(0,0){\strut{}$y/M$}}}%
      \put(3808,154){\makebox(0,0){\strut{}$x/M$}}%
    }%
    \gplgaddtomacro\gplfronttext{%
      \csname LTb\endcsname%
      \put(4857,4602){\makebox(0,0)[r]{\strut{}t=0M}}%
      \csname LTb\endcsname%
      \put(4857,4382){\makebox(0,0)[r]{\strut{}t=25M}}%
      \csname LTb\endcsname%
      \put(4857,4162){\makebox(0,0)[r]{\strut{}t=50M}}%
      \csname LTb\endcsname%
      \put(4857,3942){\makebox(0,0)[r]{\strut{}t=75M}}%
      \csname LTb\endcsname%
      \put(4857,3722){\makebox(0,0)[r]{\strut{}t=100M}}%
      \csname LTb\endcsname%
      \put(4857,3502){\makebox(0,0)[r]{\strut{}t=125M}}%
      \csname LTb\endcsname%
      \put(4857,3282){\makebox(0,0)[r]{\strut{}t=150M}}%
    }%
    \gplbacktext
    \put(0,0){\includegraphics{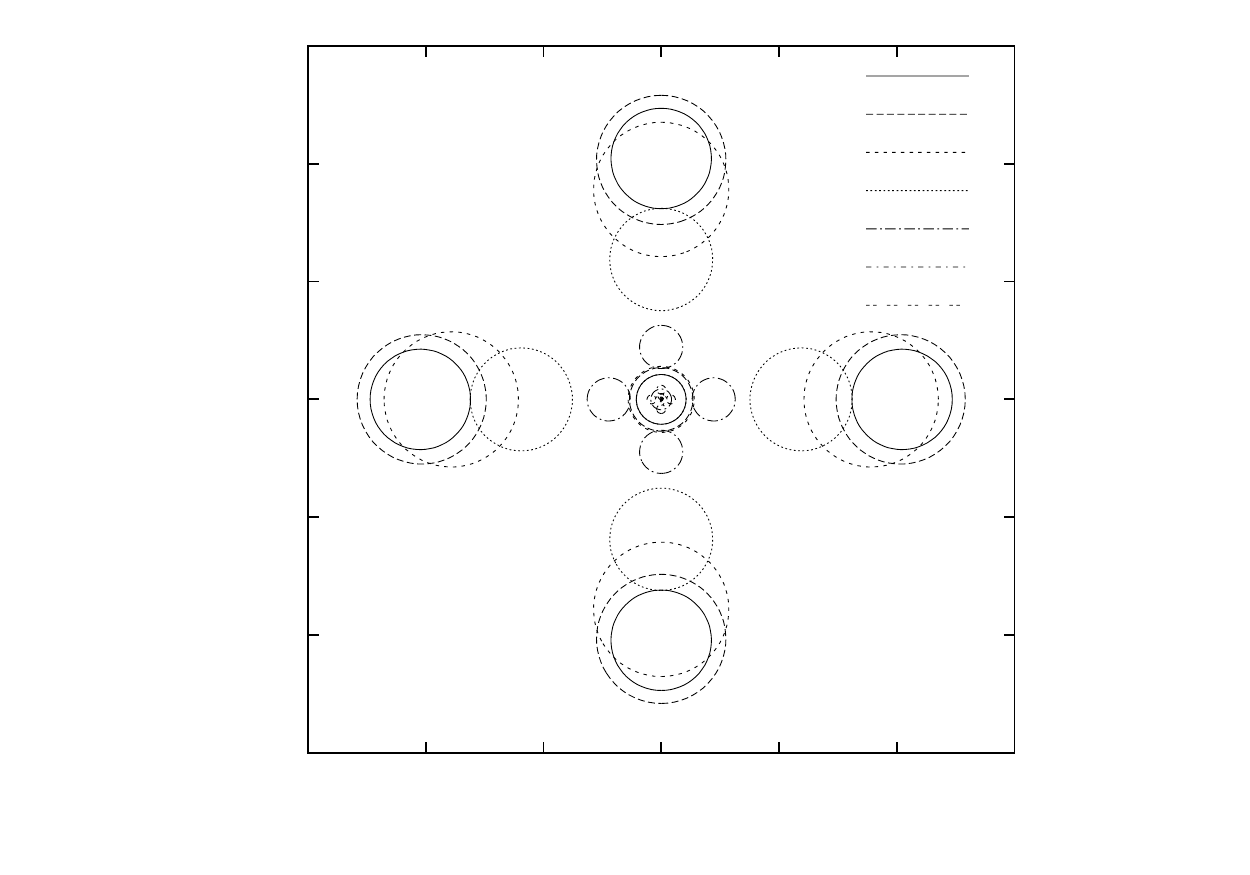}}%
    \gplfronttext
  \end{picture}%
\endgroup

\ece
\caption{Section of the eight marginal surfaces on the $z=0$ plane, at times
$t = (0,25,50,75,100,125,150)M$. The bottom plot is a zoomed-in version of
the seven central MOTSs only.\label{fig:horizons}}
\efi

\bfi
\begingroup
  \makeatletter
  \providecommand\color[2][]{%
    \GenericError{(gnuplot) \space\space\space\@spaces}{%
      Package color not loaded in conjunction with
      terminal option `colourtext'%
    }{See the gnuplot documentation for explanation.%
    }{Either use 'blacktext' in gnuplot or load the package
      color.sty in LaTeX.}%
    \renewcommand\color[2][]{}%
  }%
  \providecommand\includegraphics[2][]{%
    \GenericError{(gnuplot) \space\space\space\@spaces}{%
      Package graphicx or graphics not loaded%
    }{See the gnuplot documentation for explanation.%
    }{The gnuplot epslatex terminal needs graphicx.sty or graphics.sty.}%
    \renewcommand\includegraphics[2][]{}%
  }%
  \providecommand\rotatebox[2]{#2}%
  \@ifundefined{ifGPcolor}{%
    \newif\ifGPcolor
    \GPcolorfalse
  }{}%
  \@ifundefined{ifGPblacktext}{%
    \newif\ifGPblacktext
    \GPblacktexttrue
  }{}%
  \let\gplgaddtomacro\g@addto@macro
  \gdef\gplbacktext{}%
  \gdef\gplfronttext{}%
  \makeatother
  \ifGPblacktext
    \def\colorrgb#1{}%
    \def\colorgray#1{}%
  \else
    \ifGPcolor
      \def\colorrgb#1{\color[rgb]{#1}}%
      \def\colorgray#1{\color[gray]{#1}}%
      \expandafter\def\csname LTw\endcsname{\color{white}}%
      \expandafter\def\csname LTb\endcsname{\color{black}}%
      \expandafter\def\csname LTa\endcsname{\color{black}}%
      \expandafter\def\csname LT0\endcsname{\color[rgb]{1,0,0}}%
      \expandafter\def\csname LT1\endcsname{\color[rgb]{0,1,0}}%
      \expandafter\def\csname LT2\endcsname{\color[rgb]{0,0,1}}%
      \expandafter\def\csname LT3\endcsname{\color[rgb]{1,0,1}}%
      \expandafter\def\csname LT4\endcsname{\color[rgb]{0,1,1}}%
      \expandafter\def\csname LT5\endcsname{\color[rgb]{1,1,0}}%
      \expandafter\def\csname LT6\endcsname{\color[rgb]{0,0,0}}%
      \expandafter\def\csname LT7\endcsname{\color[rgb]{1,0.3,0}}%
      \expandafter\def\csname LT8\endcsname{\color[rgb]{0.5,0.5,0.5}}%
    \else
      \def\colorrgb#1{\color{black}}%
      \def\colorgray#1{\color[gray]{#1}}%
      \expandafter\def\csname LTw\endcsname{\color{white}}%
      \expandafter\def\csname LTb\endcsname{\color{black}}%
      \expandafter\def\csname LTa\endcsname{\color{black}}%
      \expandafter\def\csname LT0\endcsname{\color{black}}%
      \expandafter\def\csname LT1\endcsname{\color{black}}%
      \expandafter\def\csname LT2\endcsname{\color{black}}%
      \expandafter\def\csname LT3\endcsname{\color{black}}%
      \expandafter\def\csname LT4\endcsname{\color{black}}%
      \expandafter\def\csname LT5\endcsname{\color{black}}%
      \expandafter\def\csname LT6\endcsname{\color{black}}%
      \expandafter\def\csname LT7\endcsname{\color{black}}%
      \expandafter\def\csname LT8\endcsname{\color{black}}%
    \fi
  \fi
  \setlength{\unitlength}{0.0500bp}%
  \begin{picture}(7200.00,5040.00)%
    \gplgaddtomacro\gplbacktext{%
      \csname LTb\endcsname%
      \put(1839,704){\makebox(0,0)[r]{\strut{}-0.25}}%
      \put(1839,1111){\makebox(0,0)[r]{\strut{}-0.2}}%
      \put(1839,1518){\makebox(0,0)[r]{\strut{}-0.15}}%
      \put(1839,1925){\makebox(0,0)[r]{\strut{}-0.1}}%
      \put(1839,2332){\makebox(0,0)[r]{\strut{}-0.05}}%
      \put(1839,2740){\makebox(0,0)[r]{\strut{} 0}}%
      \put(1839,3147){\makebox(0,0)[r]{\strut{} 0.05}}%
      \put(1839,3554){\makebox(0,0)[r]{\strut{} 0.1}}%
      \put(1839,3961){\makebox(0,0)[r]{\strut{} 0.15}}%
      \put(1839,4368){\makebox(0,0)[r]{\strut{} 0.2}}%
      \put(1839,4775){\makebox(0,0)[r]{\strut{} 0.25}}%
      \put(1971,484){\makebox(0,0){\strut{}-0.25}}%
      \put(2378,484){\makebox(0,0){\strut{}-0.2}}%
      \put(2785,484){\makebox(0,0){\strut{}-0.15}}%
      \put(3192,484){\makebox(0,0){\strut{}-0.1}}%
      \put(3599,484){\makebox(0,0){\strut{}-0.05}}%
      \put(4007,484){\makebox(0,0){\strut{} 0}}%
      \put(4414,484){\makebox(0,0){\strut{} 0.05}}%
      \put(4821,484){\makebox(0,0){\strut{} 0.1}}%
      \put(5228,484){\makebox(0,0){\strut{} 0.15}}%
      \put(5635,484){\makebox(0,0){\strut{} 0.2}}%
      \put(6042,484){\makebox(0,0){\strut{} 0.25}}%
      \put(937,2739){\rotatebox{-270}{\makebox(0,0){\strut{}$y/M$}}}%
      \put(4006,154){\makebox(0,0){\strut{}$x/M$}}%
    }%
    \gplgaddtomacro\gplfronttext{%
    }%
    \gplbacktext
    \put(0,0){\includegraphics{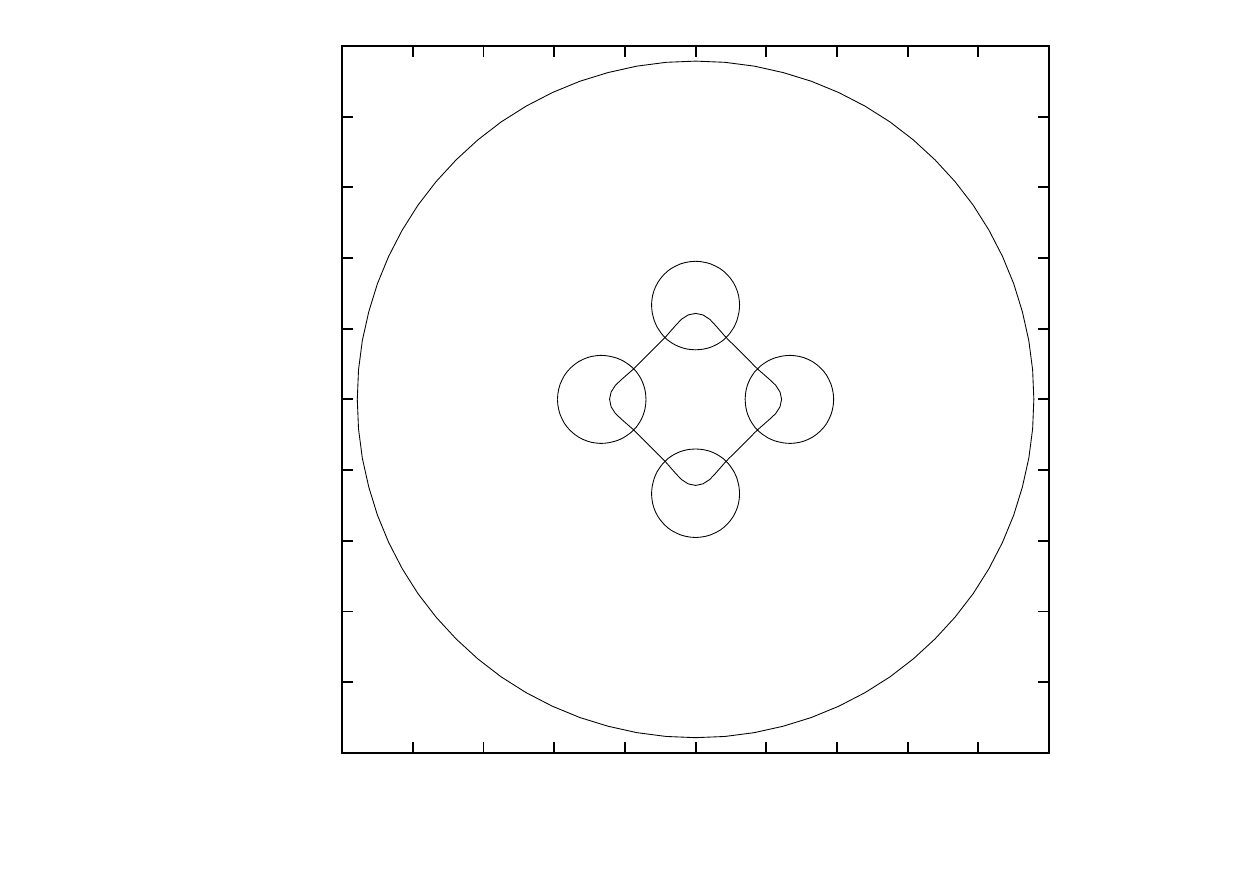}}%
    \gplfronttext
  \end{picture}%
\endgroup

\begingroup
  \makeatletter
  \providecommand\color[2][]{%
    \GenericError{(gnuplot) \space\space\space\@spaces}{%
      Package color not loaded in conjunction with
      terminal option `colourtext'%
    }{See the gnuplot documentation for explanation.%
    }{Either use 'blacktext' in gnuplot or load the package
      color.sty in LaTeX.}%
    \renewcommand\color[2][]{}%
  }%
  \providecommand\includegraphics[2][]{%
    \GenericError{(gnuplot) \space\space\space\@spaces}{%
      Package graphicx or graphics not loaded%
    }{See the gnuplot documentation for explanation.%
    }{The gnuplot epslatex terminal needs graphicx.sty or graphics.sty.}%
    \renewcommand\includegraphics[2][]{}%
  }%
  \providecommand\rotatebox[2]{#2}%
  \@ifundefined{ifGPcolor}{%
    \newif\ifGPcolor
    \GPcolorfalse
  }{}%
  \@ifundefined{ifGPblacktext}{%
    \newif\ifGPblacktext
    \GPblacktexttrue
  }{}%
  \let\gplgaddtomacro\g@addto@macro
  \gdef\gplbacktext{}%
  \gdef\gplfronttext{}%
  \makeatother
  \ifGPblacktext
    \def\colorrgb#1{}%
    \def\colorgray#1{}%
  \else
    \ifGPcolor
      \def\colorrgb#1{\color[rgb]{#1}}%
      \def\colorgray#1{\color[gray]{#1}}%
      \expandafter\def\csname LTw\endcsname{\color{white}}%
      \expandafter\def\csname LTb\endcsname{\color{black}}%
      \expandafter\def\csname LTa\endcsname{\color{black}}%
      \expandafter\def\csname LT0\endcsname{\color[rgb]{1,0,0}}%
      \expandafter\def\csname LT1\endcsname{\color[rgb]{0,1,0}}%
      \expandafter\def\csname LT2\endcsname{\color[rgb]{0,0,1}}%
      \expandafter\def\csname LT3\endcsname{\color[rgb]{1,0,1}}%
      \expandafter\def\csname LT4\endcsname{\color[rgb]{0,1,1}}%
      \expandafter\def\csname LT5\endcsname{\color[rgb]{1,1,0}}%
      \expandafter\def\csname LT6\endcsname{\color[rgb]{0,0,0}}%
      \expandafter\def\csname LT7\endcsname{\color[rgb]{1,0.3,0}}%
      \expandafter\def\csname LT8\endcsname{\color[rgb]{0.5,0.5,0.5}}%
    \else
      \def\colorrgb#1{\color{black}}%
      \def\colorgray#1{\color[gray]{#1}}%
      \expandafter\def\csname LTw\endcsname{\color{white}}%
      \expandafter\def\csname LTb\endcsname{\color{black}}%
      \expandafter\def\csname LTa\endcsname{\color{black}}%
      \expandafter\def\csname LT0\endcsname{\color{black}}%
      \expandafter\def\csname LT1\endcsname{\color{black}}%
      \expandafter\def\csname LT2\endcsname{\color{black}}%
      \expandafter\def\csname LT3\endcsname{\color{black}}%
      \expandafter\def\csname LT4\endcsname{\color{black}}%
      \expandafter\def\csname LT5\endcsname{\color{black}}%
      \expandafter\def\csname LT6\endcsname{\color{black}}%
      \expandafter\def\csname LT7\endcsname{\color{black}}%
      \expandafter\def\csname LT8\endcsname{\color{black}}%
    \fi
  \fi
  \setlength{\unitlength}{0.0500bp}%
  \begin{picture}(7200.00,5040.00)%
    \gplgaddtomacro\gplbacktext{%
      \csname LTb\endcsname%
      \put(1839,704){\makebox(0,0)[r]{\strut{}-0.03}}%
      \put(1839,1383){\makebox(0,0)[r]{\strut{}-0.02}}%
      \put(1839,2061){\makebox(0,0)[r]{\strut{}-0.01}}%
      \put(1839,2740){\makebox(0,0)[r]{\strut{} 0}}%
      \put(1839,3418){\makebox(0,0)[r]{\strut{} 0.01}}%
      \put(1839,4097){\makebox(0,0)[r]{\strut{} 0.02}}%
      \put(1839,4775){\makebox(0,0)[r]{\strut{} 0.03}}%
      \put(1971,484){\makebox(0,0){\strut{}-0.03}}%
      \put(2650,484){\makebox(0,0){\strut{}-0.02}}%
      \put(3328,484){\makebox(0,0){\strut{}-0.01}}%
      \put(4007,484){\makebox(0,0){\strut{} 0}}%
      \put(4685,484){\makebox(0,0){\strut{} 0.01}}%
      \put(5364,484){\makebox(0,0){\strut{} 0.02}}%
      \put(6042,484){\makebox(0,0){\strut{} 0.03}}%
      \put(937,2739){\rotatebox{-270}{\makebox(0,0){\strut{}$y/M$}}}%
      \put(4006,154){\makebox(0,0){\strut{}$x/M$}}%
    }%
    \gplgaddtomacro\gplfronttext{%
    }%
    \gplbacktext
    \put(0,0){\includegraphics{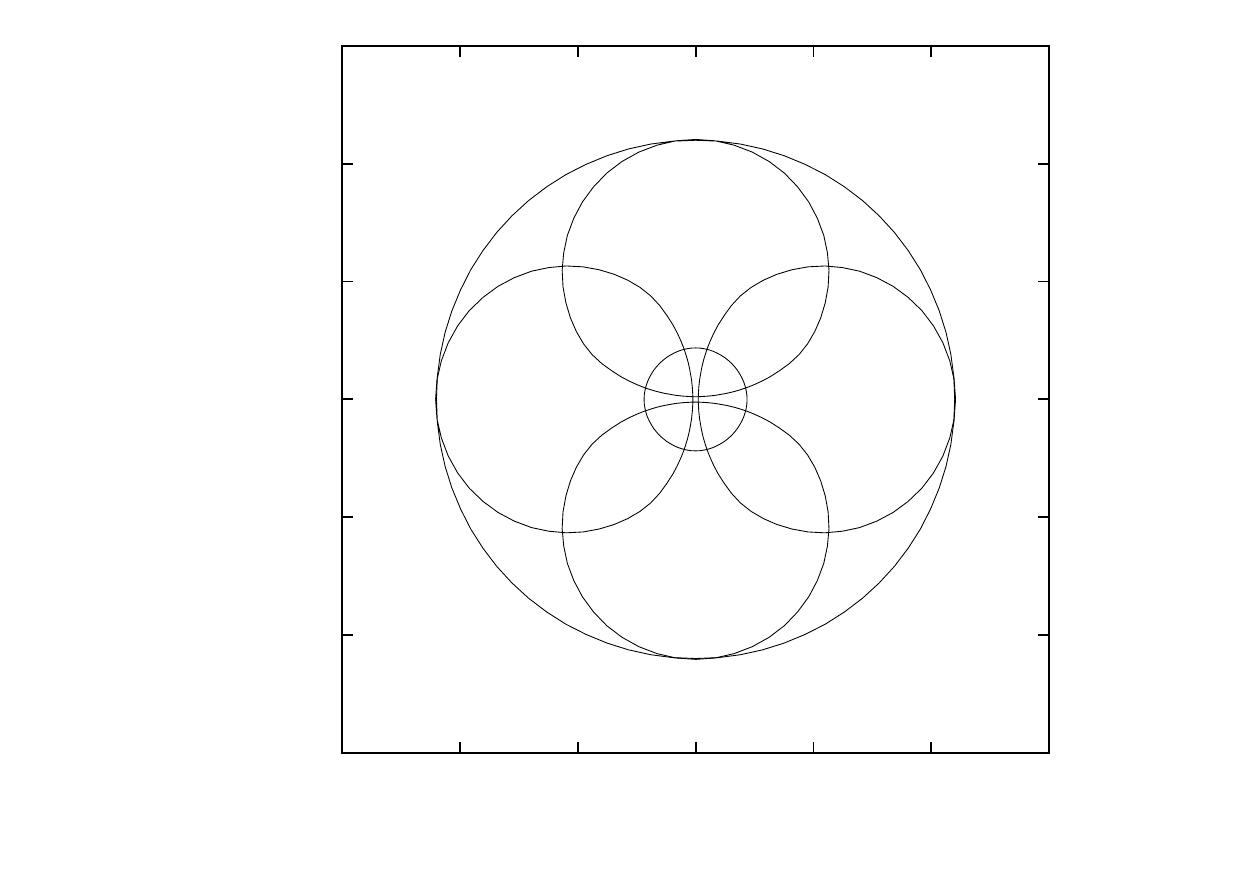}}%
    \gplfronttext
  \end{picture}%
\endgroup

\caption{Section of the eight marginal surfaces on the $z=0$ plane, on the
spatial slice of the first surface merger, corresponding to coordinate time
$t=128M$ (top) and on that of the second merger at $t=167M$ (bottom).\label{fig:merger}}
\efi

\bfi
\bce
\begingroup
  \makeatletter
  \providecommand\color[2][]{%
    \GenericError{(gnuplot) \space\space\space\@spaces}{%
      Package color not loaded in conjunction with
      terminal option `colourtext'%
    }{See the gnuplot documentation for explanation.%
    }{Either use 'blacktext' in gnuplot or load the package
      color.sty in LaTeX.}%
    \renewcommand\color[2][]{}%
  }%
  \providecommand\includegraphics[2][]{%
    \GenericError{(gnuplot) \space\space\space\@spaces}{%
      Package graphicx or graphics not loaded%
    }{See the gnuplot documentation for explanation.%
    }{The gnuplot epslatex terminal needs graphicx.sty or graphics.sty.}%
    \renewcommand\includegraphics[2][]{}%
  }%
  \providecommand\rotatebox[2]{#2}%
  \@ifundefined{ifGPcolor}{%
    \newif\ifGPcolor
    \GPcolorfalse
  }{}%
  \@ifundefined{ifGPblacktext}{%
    \newif\ifGPblacktext
    \GPblacktexttrue
  }{}%
  \let\gplgaddtomacro\g@addto@macro
  \gdef\gplbacktext{}%
  \gdef\gplfronttext{}%
  \makeatother
  \ifGPblacktext
    \def\colorrgb#1{}%
    \def\colorgray#1{}%
  \else
    \ifGPcolor
      \def\colorrgb#1{\color[rgb]{#1}}%
      \def\colorgray#1{\color[gray]{#1}}%
      \expandafter\def\csname LTw\endcsname{\color{white}}%
      \expandafter\def\csname LTb\endcsname{\color{black}}%
      \expandafter\def\csname LTa\endcsname{\color{black}}%
      \expandafter\def\csname LT0\endcsname{\color[rgb]{1,0,0}}%
      \expandafter\def\csname LT1\endcsname{\color[rgb]{0,1,0}}%
      \expandafter\def\csname LT2\endcsname{\color[rgb]{0,0,1}}%
      \expandafter\def\csname LT3\endcsname{\color[rgb]{1,0,1}}%
      \expandafter\def\csname LT4\endcsname{\color[rgb]{0,1,1}}%
      \expandafter\def\csname LT5\endcsname{\color[rgb]{1,1,0}}%
      \expandafter\def\csname LT6\endcsname{\color[rgb]{0,0,0}}%
      \expandafter\def\csname LT7\endcsname{\color[rgb]{1,0.3,0}}%
      \expandafter\def\csname LT8\endcsname{\color[rgb]{0.5,0.5,0.5}}%
    \else
      \def\colorrgb#1{\color{black}}%
      \def\colorgray#1{\color[gray]{#1}}%
      \expandafter\def\csname LTw\endcsname{\color{white}}%
      \expandafter\def\csname LTb\endcsname{\color{black}}%
      \expandafter\def\csname LTa\endcsname{\color{black}}%
      \expandafter\def\csname LT0\endcsname{\color{black}}%
      \expandafter\def\csname LT1\endcsname{\color{black}}%
      \expandafter\def\csname LT2\endcsname{\color{black}}%
      \expandafter\def\csname LT3\endcsname{\color{black}}%
      \expandafter\def\csname LT4\endcsname{\color{black}}%
      \expandafter\def\csname LT5\endcsname{\color{black}}%
      \expandafter\def\csname LT6\endcsname{\color{black}}%
      \expandafter\def\csname LT7\endcsname{\color{black}}%
      \expandafter\def\csname LT8\endcsname{\color{black}}%
    \fi
  \fi
  \setlength{\unitlength}{0.0500bp}%
  \begin{picture}(5760.00,4032.00)%
    \gplgaddtomacro\gplbacktext{%
      \csname LTb\endcsname%
      \put(1078,704){\makebox(0,0)[r]{\strut{} 37.5}}%
      \put(1078,1142){\makebox(0,0)[r]{\strut{} 38}}%
      \put(1078,1579){\makebox(0,0)[r]{\strut{} 38.5}}%
      \put(1078,2017){\makebox(0,0)[r]{\strut{} 39}}%
      \put(1078,2454){\makebox(0,0)[r]{\strut{} 39.5}}%
      \put(1078,2892){\makebox(0,0)[r]{\strut{} 40}}%
      \put(1078,3329){\makebox(0,0)[r]{\strut{} 40.5}}%
      \put(1078,3767){\makebox(0,0)[r]{\strut{} 41}}%
      \put(1210,484){\makebox(0,0){\strut{} 0}}%
      \put(1864,484){\makebox(0,0){\strut{} 20}}%
      \put(2518,484){\makebox(0,0){\strut{} 40}}%
      \put(3172,484){\makebox(0,0){\strut{} 60}}%
      \put(3826,484){\makebox(0,0){\strut{} 80}}%
      \put(4480,484){\makebox(0,0){\strut{} 100}}%
      \put(5134,484){\makebox(0,0){\strut{} 120}}%
      \put(176,2235){\rotatebox{-270}{\makebox(0,0){\strut{}Mass}}}%
      \put(3286,154){\makebox(0,0){\strut{}$t/M$}}%
    }%
    \gplgaddtomacro\gplfronttext{%
      \csname LTb\endcsname%
      \put(4376,3594){\makebox(0,0)[r]{\strut{}Horizon 1}}%
      \csname LTb\endcsname%
      \put(4376,3374){\makebox(0,0)[r]{\strut{}Horizon 2}}%
      \csname LTb\endcsname%
      \put(4376,3154){\makebox(0,0)[r]{\strut{}Horizon 3}}%
    }%
    \gplbacktext
    \put(0,0){\includegraphics{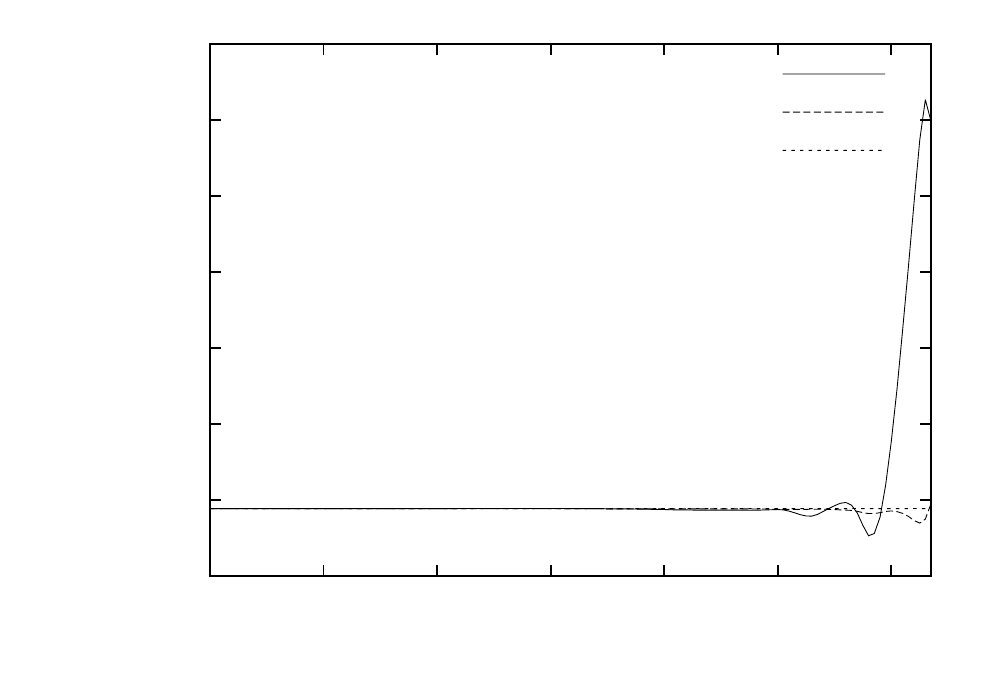}}%
    \gplfronttext
  \end{picture}%
\endgroup

\caption{Masses of the three marginal surfaces of the black holes at (0,0,0), (2,2,2) and 
infinity, as a function of coordinate time.\label{fig:mass}}
\ece
\efi

\bfi
\bce
\begingroup
  \makeatletter
  \providecommand\color[2][]{%
    \GenericError{(gnuplot) \space\space\space\@spaces}{%
      Package color not loaded in conjunction with
      terminal option `colourtext'%
    }{See the gnuplot documentation for explanation.%
    }{Either use 'blacktext' in gnuplot or load the package
      color.sty in LaTeX.}%
    \renewcommand\color[2][]{}%
  }%
  \providecommand\includegraphics[2][]{%
    \GenericError{(gnuplot) \space\space\space\@spaces}{%
      Package graphicx or graphics not loaded%
    }{See the gnuplot documentation for explanation.%
    }{The gnuplot epslatex terminal needs graphicx.sty or graphics.sty.}%
    \renewcommand\includegraphics[2][]{}%
  }%
  \providecommand\rotatebox[2]{#2}%
  \@ifundefined{ifGPcolor}{%
    \newif\ifGPcolor
    \GPcolorfalse
  }{}%
  \@ifundefined{ifGPblacktext}{%
    \newif\ifGPblacktext
    \GPblacktexttrue
  }{}%
  \let\gplgaddtomacro\g@addto@macro
  \gdef\gplbacktext{}%
  \gdef\gplfronttext{}%
  \makeatother
  \ifGPblacktext
    \def\colorrgb#1{}%
    \def\colorgray#1{}%
  \else
    \ifGPcolor
      \def\colorrgb#1{\color[rgb]{#1}}%
      \def\colorgray#1{\color[gray]{#1}}%
      \expandafter\def\csname LTw\endcsname{\color{white}}%
      \expandafter\def\csname LTb\endcsname{\color{black}}%
      \expandafter\def\csname LTa\endcsname{\color{black}}%
      \expandafter\def\csname LT0\endcsname{\color[rgb]{1,0,0}}%
      \expandafter\def\csname LT1\endcsname{\color[rgb]{0,1,0}}%
      \expandafter\def\csname LT2\endcsname{\color[rgb]{0,0,1}}%
      \expandafter\def\csname LT3\endcsname{\color[rgb]{1,0,1}}%
      \expandafter\def\csname LT4\endcsname{\color[rgb]{0,1,1}}%
      \expandafter\def\csname LT5\endcsname{\color[rgb]{1,1,0}}%
      \expandafter\def\csname LT6\endcsname{\color[rgb]{0,0,0}}%
      \expandafter\def\csname LT7\endcsname{\color[rgb]{1,0.3,0}}%
      \expandafter\def\csname LT8\endcsname{\color[rgb]{0.5,0.5,0.5}}%
    \else
      \def\colorrgb#1{\color{black}}%
      \def\colorgray#1{\color[gray]{#1}}%
      \expandafter\def\csname LTw\endcsname{\color{white}}%
      \expandafter\def\csname LTb\endcsname{\color{black}}%
      \expandafter\def\csname LTa\endcsname{\color{black}}%
      \expandafter\def\csname LT0\endcsname{\color{black}}%
      \expandafter\def\csname LT1\endcsname{\color{black}}%
      \expandafter\def\csname LT2\endcsname{\color{black}}%
      \expandafter\def\csname LT3\endcsname{\color{black}}%
      \expandafter\def\csname LT4\endcsname{\color{black}}%
      \expandafter\def\csname LT5\endcsname{\color{black}}%
      \expandafter\def\csname LT6\endcsname{\color{black}}%
      \expandafter\def\csname LT7\endcsname{\color{black}}%
      \expandafter\def\csname LT8\endcsname{\color{black}}%
    \fi
  \fi
  \setlength{\unitlength}{0.0500bp}%
  \begin{picture}(5760.00,4032.00)%
    \gplgaddtomacro\gplbacktext{%
      \csname LTb\endcsname%
      \put(1078,704){\makebox(0,0)[r]{\strut{} 0.01}}%
      \put(1078,1470){\makebox(0,0)[r]{\strut{} 0.1}}%
      \put(1078,2236){\makebox(0,0)[r]{\strut{} 1}}%
      \put(1078,3001){\makebox(0,0)[r]{\strut{} 10}}%
      \put(1078,3767){\makebox(0,0)[r]{\strut{} 100}}%
      \put(1210,484){\makebox(0,0){\strut{} 0}}%
      \put(1864,484){\makebox(0,0){\strut{} 20}}%
      \put(2518,484){\makebox(0,0){\strut{} 40}}%
      \put(3172,484){\makebox(0,0){\strut{} 60}}%
      \put(3826,484){\makebox(0,0){\strut{} 80}}%
      \put(4480,484){\makebox(0,0){\strut{} 100}}%
      \put(5134,484){\makebox(0,0){\strut{} 120}}%
      \put(176,2235){\rotatebox{-270}{\makebox(0,0){\strut{}Mean coordinate radius}}}%
      \put(3286,154){\makebox(0,0){\strut{}$t/M$}}%
    }%
    \gplgaddtomacro\gplfronttext{%
      \csname LTb\endcsname%
      \put(4376,3594){\makebox(0,0)[r]{\strut{}Horizon 1}}%
      \csname LTb\endcsname%
      \put(4376,3374){\makebox(0,0)[r]{\strut{}Horizon 2}}%
      \csname LTb\endcsname%
      \put(4376,3154){\makebox(0,0)[r]{\strut{}Horizon 3}}%
    }%
    \gplbacktext
    \put(0,0){\includegraphics{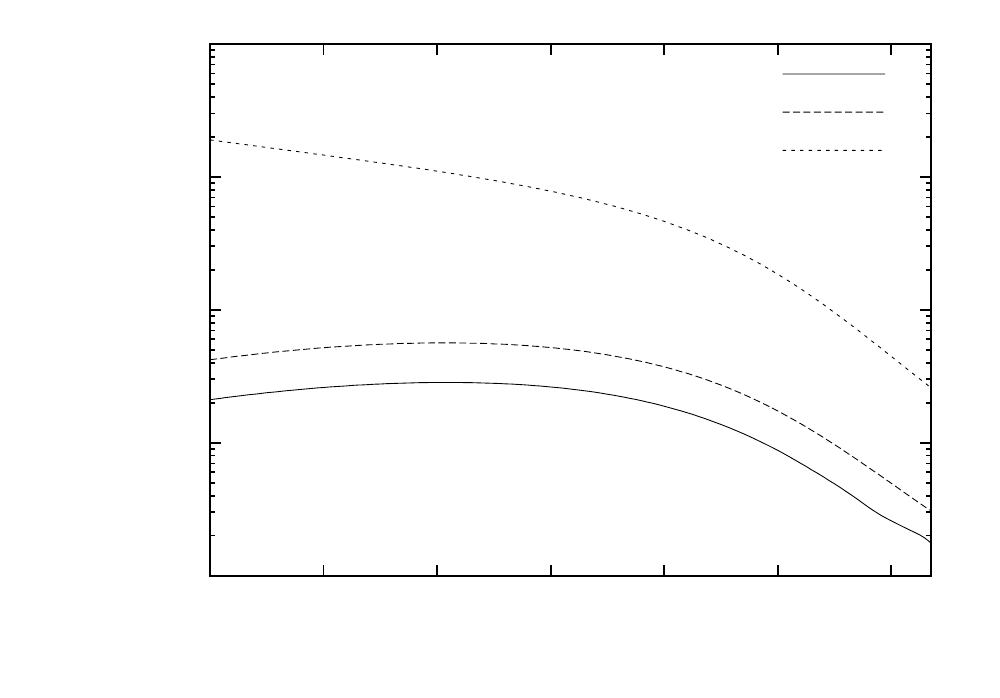}}%
    \gplfronttext
  \end{picture}%
\endgroup

\caption{Mean coordinate radii of the three marginal surfaces of the black holes at (0,0,0), (2,2,2) and 
infinity, as a function of coordinate time.\label{fig:mradius}}
\ece
\efi

From the cosmological standpoint, we are particularly interested in the scaling
of lengths with proper time. There are at least two candidate quantities for
measuring the scaling of distances in this system: the (minimal) proper distance between
near-neighbour surfaces and the proper length of each cell's edges. Note that these 
estimates need not agree with each other, since the expansion rate may well be different at different points. 

In order to calculate proper lengths as functions of proper time,
we restrict our attention to the 1+1 subspace spanned by a reprentative curve in
time, and obtain the proper time and the $x$-coordinate of gaussian observers with
the following relations:
\bea
\label{eq:tgauss}
&&\tau(t,x) =\int_0^t \alpha(t',x) {\rm d} t'\\
\label{eq:xgauss}
&&x_{\rm g}(t,x_{\rm init}) = x_{\rm g}(t-\Delta t,x_{\rm init})-\int^t_{t-\Delta t} \beta^x(t',x_{\rm g}(t-\Delta t,x_{\rm init})) {\rm d} t'
\eea
where $x_{\rm init}$ is the location of the observer at $t=0$.

The proper distance between marginal surfaces, as a function of proper time, is then given by:
\bea
\label{eq:Dh}
\fl
D(\tau)=\int_{\gamma_\tau} \left [  (-\alpha^2(\tau,\ell)+\beta^2(\tau,\ell)) (\partial_\ell t)^2 
                                   + \beta_i(\tau,\ell) \partial_\ell t \partial_\ell x^i 
                                   + \gamma_{ij}(\tau,\ell) \partial_\ell x^i \partial_\ell x^j \right ]^{1/2} {\rm d} \ell && \nonumber \\
\eea
where $\gamma_\tau$ is the shortest constant-$\tau$ geodesic, parametrized by $\ell$, connecting two
surfaces. We measure this quantity for the two outer-trapped surfaces initially at (0,0,0)
and (2,0,0), the geodesic lying on the $x$-axis for symmetry reasons. This quantity
is plotted in Figure~\ref{fig:pd}. 

\bfi
\bce
\begingroup
  \makeatletter
  \providecommand\color[2][]{%
    \GenericError{(gnuplot) \space\space\space\@spaces}{%
      Package color not loaded in conjunction with
      terminal option `colourtext'%
    }{See the gnuplot documentation for explanation.%
    }{Either use 'blacktext' in gnuplot or load the package
      color.sty in LaTeX.}%
    \renewcommand\color[2][]{}%
  }%
  \providecommand\includegraphics[2][]{%
    \GenericError{(gnuplot) \space\space\space\@spaces}{%
      Package graphicx or graphics not loaded%
    }{See the gnuplot documentation for explanation.%
    }{The gnuplot epslatex terminal needs graphicx.sty or graphics.sty.}%
    \renewcommand\includegraphics[2][]{}%
  }%
  \providecommand\rotatebox[2]{#2}%
  \@ifundefined{ifGPcolor}{%
    \newif\ifGPcolor
    \GPcolorfalse
  }{}%
  \@ifundefined{ifGPblacktext}{%
    \newif\ifGPblacktext
    \GPblacktexttrue
  }{}%
  \let\gplgaddtomacro\g@addto@macro
  \gdef\gplbacktext{}%
  \gdef\gplfronttext{}%
  \makeatother
  \ifGPblacktext
    \def\colorrgb#1{}%
    \def\colorgray#1{}%
  \else
    \ifGPcolor
      \def\colorrgb#1{\color[rgb]{#1}}%
      \def\colorgray#1{\color[gray]{#1}}%
      \expandafter\def\csname LTw\endcsname{\color{white}}%
      \expandafter\def\csname LTb\endcsname{\color{black}}%
      \expandafter\def\csname LTa\endcsname{\color{black}}%
      \expandafter\def\csname LT0\endcsname{\color[rgb]{1,0,0}}%
      \expandafter\def\csname LT1\endcsname{\color[rgb]{0,1,0}}%
      \expandafter\def\csname LT2\endcsname{\color[rgb]{0,0,1}}%
      \expandafter\def\csname LT3\endcsname{\color[rgb]{1,0,1}}%
      \expandafter\def\csname LT4\endcsname{\color[rgb]{0,1,1}}%
      \expandafter\def\csname LT5\endcsname{\color[rgb]{1,1,0}}%
      \expandafter\def\csname LT6\endcsname{\color[rgb]{0,0,0}}%
      \expandafter\def\csname LT7\endcsname{\color[rgb]{1,0.3,0}}%
      \expandafter\def\csname LT8\endcsname{\color[rgb]{0.5,0.5,0.5}}%
    \else
      \def\colorrgb#1{\color{black}}%
      \def\colorgray#1{\color[gray]{#1}}%
      \expandafter\def\csname LTw\endcsname{\color{white}}%
      \expandafter\def\csname LTb\endcsname{\color{black}}%
      \expandafter\def\csname LTa\endcsname{\color{black}}%
      \expandafter\def\csname LT0\endcsname{\color{black}}%
      \expandafter\def\csname LT1\endcsname{\color{black}}%
      \expandafter\def\csname LT2\endcsname{\color{black}}%
      \expandafter\def\csname LT3\endcsname{\color{black}}%
      \expandafter\def\csname LT4\endcsname{\color{black}}%
      \expandafter\def\csname LT5\endcsname{\color{black}}%
      \expandafter\def\csname LT6\endcsname{\color{black}}%
      \expandafter\def\csname LT7\endcsname{\color{black}}%
      \expandafter\def\csname LT8\endcsname{\color{black}}%
    \fi
  \fi
  \setlength{\unitlength}{0.0500bp}%
  \begin{picture}(5760.00,4032.00)%
    \gplgaddtomacro\gplbacktext{%
      \csname LTb\endcsname%
      \put(726,704){\makebox(0,0)[r]{\strut{} 0.5}}%
      \put(726,1087){\makebox(0,0)[r]{\strut{} 0.6}}%
      \put(726,1470){\makebox(0,0)[r]{\strut{} 0.7}}%
      \put(726,1853){\makebox(0,0)[r]{\strut{} 0.8}}%
      \put(726,2236){\makebox(0,0)[r]{\strut{} 0.9}}%
      \put(726,2618){\makebox(0,0)[r]{\strut{} 1}}%
      \put(726,3001){\makebox(0,0)[r]{\strut{} 1.1}}%
      \put(726,3384){\makebox(0,0)[r]{\strut{} 1.2}}%
      \put(726,3767){\makebox(0,0)[r]{\strut{} 1.3}}%
      \put(858,484){\makebox(0,0){\strut{} 0}}%
      \put(1759,484){\makebox(0,0){\strut{} 20}}%
      \put(2660,484){\makebox(0,0){\strut{} 40}}%
      \put(3561,484){\makebox(0,0){\strut{} 60}}%
      \put(4462,484){\makebox(0,0){\strut{} 80}}%
      \put(5363,484){\makebox(0,0){\strut{} 100}}%
      \put(3110,154){\makebox(0,0){\strut{}$\tau/M$}}%
    }%
    \gplgaddtomacro\gplfronttext{%
      \csname LTb\endcsname%
      \put(4376,3594){\makebox(0,0)[r]{\strut{}Horizon distance}}%
      \csname LTb\endcsname%
      \put(4376,3374){\makebox(0,0)[r]{\strut{}Edge length}}%
      \csname LTb\endcsname%
      \put(4376,3154){\makebox(0,0)[r]{\strut{}FLRW}}%
    }%
    \gplbacktext
    \put(0,0){\includegraphics{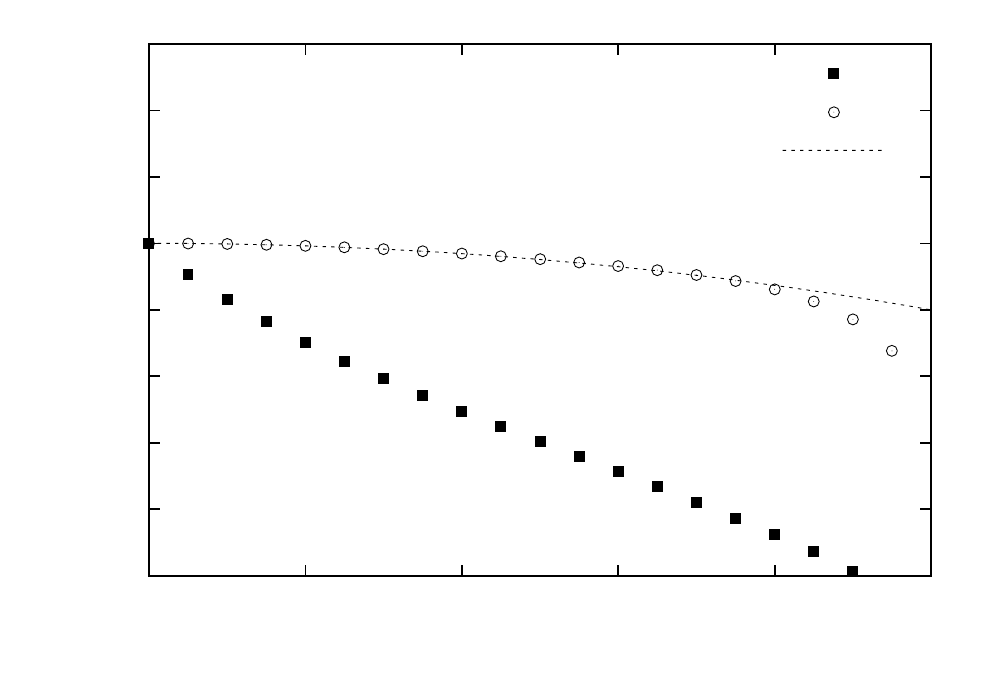}}%
    \gplfronttext
  \end{picture}%
\endgroup

\caption{Several measures of scaling in the eight-black-hole universe, as functions of
proper time $\tau$, plotted against a possible identification of the corresponding FLRW model
(see section~\ref{sec:contrast} for details). All the quantities have been renormalized to their respective 
values at $\tau=0$.\label{fig:pd}}
\ece
\efi

Similarly, the proper length of a lattice edge is given by equation (\ref{eq:Dh}),
where now $\gamma_\tau$ is the constant-$\tau$ geodesic, parametrized by $\ell$, connecting the two vertices.
It is easy to work out the initial shape of the cells, illustrated in Figure~\ref{fig:proj}.
The initial locations of the 16 vertices are given by ($\pm 2/3$,$\pm 2/3$,$\pm 2/3$) and
($\pm 2$,$\pm 2$,$\pm 2$). For simplicity, we choose to focus on the edge connecting 
(2/3,2/3,2/3) to (2,2,2), which, for symmetry reasons, always lies along the $x=y=z$ diagonal.
The edge's proper length as a function of proper time is also shown in Figure~\ref{fig:pd}.

For reference, the relative spatial and temporal scales of the system are illustrated 
in Figures~\ref{fig:vert} and~\ref{fig:numcoords}.
The numerical locations of the two vertices during the evolution is shown in Figure~\ref{fig:vert},
along with a few other representative points on the geodesic and the constant-$\tau$ lines.
In Figure~\ref{fig:numcoords}, we also show the span of the numerical coordinates of
the cell edge in 
$(\tau,x_{\rm g})$ space: this illustrates how the gauge condition adopted in this 
simulation freezes the evolution around $\tau \lesssim 150M$, preventing us from observing the 
system's behavior after this time. In Figure~\ref{fig:vert}, we also
plot the intersection of the marginal surfaces surrounding the black holes at the origin and at
infinity with the 
$x=y=z$ diagonal; this illustrates that, after $t \sim 120M$, the inner vertex is quite
close to the MOTS of the black hole at the origin: by this time, we can expect finite-size
effects to play a significant role in the scaling of lengths. Additionally, we show in~\ref{app:res} that the numerical
error quickly degrades after $\tau \sim 80M$. Based on these considerations, we only show the 
scaling up to proper times of about $100M$ in Figure~\ref{fig:pd}.

\bfi
\bce
\begingroup
  \makeatletter
  \providecommand\color[2][]{%
    \GenericError{(gnuplot) \space\space\space\@spaces}{%
      Package color not loaded in conjunction with
      terminal option `colourtext'%
    }{See the gnuplot documentation for explanation.%
    }{Either use 'blacktext' in gnuplot or load the package
      color.sty in LaTeX.}%
    \renewcommand\color[2][]{}%
  }%
  \providecommand\includegraphics[2][]{%
    \GenericError{(gnuplot) \space\space\space\@spaces}{%
      Package graphicx or graphics not loaded%
    }{See the gnuplot documentation for explanation.%
    }{The gnuplot epslatex terminal needs graphicx.sty or graphics.sty.}%
    \renewcommand\includegraphics[2][]{}%
  }%
  \providecommand\rotatebox[2]{#2}%
  \@ifundefined{ifGPcolor}{%
    \newif\ifGPcolor
    \GPcolorfalse
  }{}%
  \@ifundefined{ifGPblacktext}{%
    \newif\ifGPblacktext
    \GPblacktexttrue
  }{}%
  \let\gplgaddtomacro\g@addto@macro
  \gdef\gplbacktext{}%
  \gdef\gplfronttext{}%
  \makeatother
  \ifGPblacktext
    \def\colorrgb#1{}%
    \def\colorgray#1{}%
  \else
    \ifGPcolor
      \def\colorrgb#1{\color[rgb]{#1}}%
      \def\colorgray#1{\color[gray]{#1}}%
      \expandafter\def\csname LTw\endcsname{\color{white}}%
      \expandafter\def\csname LTb\endcsname{\color{black}}%
      \expandafter\def\csname LTa\endcsname{\color{black}}%
      \expandafter\def\csname LT0\endcsname{\color[rgb]{1,0,0}}%
      \expandafter\def\csname LT1\endcsname{\color[rgb]{0,1,0}}%
      \expandafter\def\csname LT2\endcsname{\color[rgb]{0,0,1}}%
      \expandafter\def\csname LT3\endcsname{\color[rgb]{1,0,1}}%
      \expandafter\def\csname LT4\endcsname{\color[rgb]{0,1,1}}%
      \expandafter\def\csname LT5\endcsname{\color[rgb]{1,1,0}}%
      \expandafter\def\csname LT6\endcsname{\color[rgb]{0,0,0}}%
      \expandafter\def\csname LT7\endcsname{\color[rgb]{1,0.3,0}}%
      \expandafter\def\csname LT8\endcsname{\color[rgb]{0.5,0.5,0.5}}%
    \else
      \def\colorrgb#1{\color{black}}%
      \def\colorgray#1{\color[gray]{#1}}%
      \expandafter\def\csname LTw\endcsname{\color{white}}%
      \expandafter\def\csname LTb\endcsname{\color{black}}%
      \expandafter\def\csname LTa\endcsname{\color{black}}%
      \expandafter\def\csname LT0\endcsname{\color{black}}%
      \expandafter\def\csname LT1\endcsname{\color{black}}%
      \expandafter\def\csname LT2\endcsname{\color{black}}%
      \expandafter\def\csname LT3\endcsname{\color{black}}%
      \expandafter\def\csname LT4\endcsname{\color{black}}%
      \expandafter\def\csname LT5\endcsname{\color{black}}%
      \expandafter\def\csname LT6\endcsname{\color{black}}%
      \expandafter\def\csname LT7\endcsname{\color{black}}%
      \expandafter\def\csname LT8\endcsname{\color{black}}%
    \fi
  \fi
  \setlength{\unitlength}{0.0500bp}%
  \begin{picture}(5760.00,4032.00)%
    \gplgaddtomacro\gplbacktext{%
      \csname LTb\endcsname%
      \put(946,704){\makebox(0,0)[r]{\strut{} 0}}%
      \put(946,1010){\makebox(0,0)[r]{\strut{} 20}}%
      \put(946,1317){\makebox(0,0)[r]{\strut{} 40}}%
      \put(946,1623){\makebox(0,0)[r]{\strut{} 60}}%
      \put(946,1929){\makebox(0,0)[r]{\strut{} 80}}%
      \put(946,2236){\makebox(0,0)[r]{\strut{} 100}}%
      \put(946,2542){\makebox(0,0)[r]{\strut{} 120}}%
      \put(946,2848){\makebox(0,0)[r]{\strut{} 140}}%
      \put(946,3154){\makebox(0,0)[r]{\strut{} 160}}%
      \put(946,3461){\makebox(0,0)[r]{\strut{} 180}}%
      \put(946,3767){\makebox(0,0)[r]{\strut{} 200}}%
      \put(1078,484){\makebox(0,0){\strut{} 0.0001}}%
      \put(1792,484){\makebox(0,0){\strut{} 0.001}}%
      \put(2506,484){\makebox(0,0){\strut{} 0.01}}%
      \put(3221,484){\makebox(0,0){\strut{} 0.1}}%
      \put(3935,484){\makebox(0,0){\strut{} 1}}%
      \put(4649,484){\makebox(0,0){\strut{} 10}}%
      \put(5363,484){\makebox(0,0){\strut{} 100}}%
      \put(176,2235){\rotatebox{-270}{\makebox(0,0){\strut{}$t/M$}}}%
      \put(3220,154){\makebox(0,0){\strut{}$x/M$}}%
    }%
    \gplgaddtomacro\gplfronttext{%
    }%
    \gplbacktext
    \put(0,0){\includegraphics{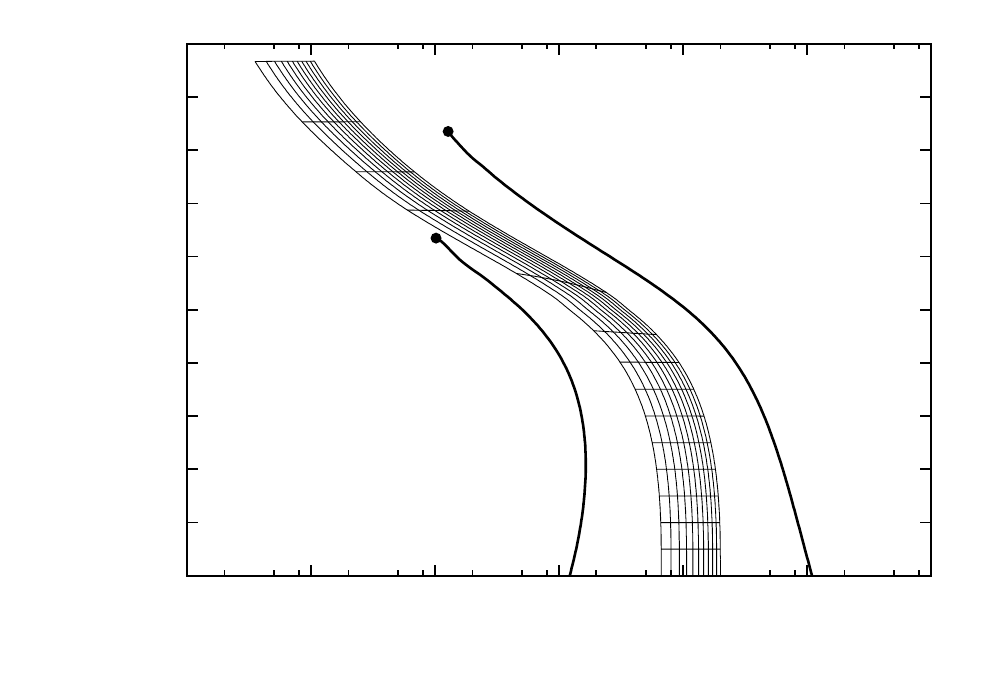}}%
    \gplfronttext
  \end{picture}%
\endgroup

\ece
\caption{Coordinate $x_{\rm g}$ of gaussian observers and constant-$\tau$ lines ($\tau=0,10, \dots , 140M$) on the $x=y=z$ diagonal, for points
initially located between the two vertices at (2/3,2/3,2/3) and (2,2,2). The thick lines represent the $x$-coordinate 
of the intersections of the marginal surfaces at (0,0,0) (left) and at infinity (right) with the diagonal, up until the mergers at $t=128M$
and $t=167M$, respectively.
\label{fig:vert}}
\efi

\bfi
\bce
\begingroup
  \makeatletter
  \providecommand\color[2][]{%
    \GenericError{(gnuplot) \space\space\space\@spaces}{%
      Package color not loaded in conjunction with
      terminal option `colourtext'%
    }{See the gnuplot documentation for explanation.%
    }{Either use 'blacktext' in gnuplot or load the package
      color.sty in LaTeX.}%
    \renewcommand\color[2][]{}%
  }%
  \providecommand\includegraphics[2][]{%
    \GenericError{(gnuplot) \space\space\space\@spaces}{%
      Package graphicx or graphics not loaded%
    }{See the gnuplot documentation for explanation.%
    }{The gnuplot epslatex terminal needs graphicx.sty or graphics.sty.}%
    \renewcommand\includegraphics[2][]{}%
  }%
  \providecommand\rotatebox[2]{#2}%
  \@ifundefined{ifGPcolor}{%
    \newif\ifGPcolor
    \GPcolorfalse
  }{}%
  \@ifundefined{ifGPblacktext}{%
    \newif\ifGPblacktext
    \GPblacktexttrue
  }{}%
  \let\gplgaddtomacro\g@addto@macro
  \gdef\gplbacktext{}%
  \gdef\gplfronttext{}%
  \makeatother
  \ifGPblacktext
    \def\colorrgb#1{}%
    \def\colorgray#1{}%
  \else
    \ifGPcolor
      \def\colorrgb#1{\color[rgb]{#1}}%
      \def\colorgray#1{\color[gray]{#1}}%
      \expandafter\def\csname LTw\endcsname{\color{white}}%
      \expandafter\def\csname LTb\endcsname{\color{black}}%
      \expandafter\def\csname LTa\endcsname{\color{black}}%
      \expandafter\def\csname LT0\endcsname{\color[rgb]{1,0,0}}%
      \expandafter\def\csname LT1\endcsname{\color[rgb]{0,1,0}}%
      \expandafter\def\csname LT2\endcsname{\color[rgb]{0,0,1}}%
      \expandafter\def\csname LT3\endcsname{\color[rgb]{1,0,1}}%
      \expandafter\def\csname LT4\endcsname{\color[rgb]{0,1,1}}%
      \expandafter\def\csname LT5\endcsname{\color[rgb]{1,1,0}}%
      \expandafter\def\csname LT6\endcsname{\color[rgb]{0,0,0}}%
      \expandafter\def\csname LT7\endcsname{\color[rgb]{1,0.3,0}}%
      \expandafter\def\csname LT8\endcsname{\color[rgb]{0.5,0.5,0.5}}%
    \else
      \def\colorrgb#1{\color{black}}%
      \def\colorgray#1{\color[gray]{#1}}%
      \expandafter\def\csname LTw\endcsname{\color{white}}%
      \expandafter\def\csname LTb\endcsname{\color{black}}%
      \expandafter\def\csname LTa\endcsname{\color{black}}%
      \expandafter\def\csname LT0\endcsname{\color{black}}%
      \expandafter\def\csname LT1\endcsname{\color{black}}%
      \expandafter\def\csname LT2\endcsname{\color{black}}%
      \expandafter\def\csname LT3\endcsname{\color{black}}%
      \expandafter\def\csname LT4\endcsname{\color{black}}%
      \expandafter\def\csname LT5\endcsname{\color{black}}%
      \expandafter\def\csname LT6\endcsname{\color{black}}%
      \expandafter\def\csname LT7\endcsname{\color{black}}%
      \expandafter\def\csname LT8\endcsname{\color{black}}%
    \fi
  \fi
  \setlength{\unitlength}{0.0500bp}%
  \begin{picture}(5760.00,4032.00)%
    \gplgaddtomacro\gplbacktext{%
      \csname LTb\endcsname%
      \put(946,704){\makebox(0,0)[r]{\strut{} 0}}%
      \put(946,1087){\makebox(0,0)[r]{\strut{} 20}}%
      \put(946,1470){\makebox(0,0)[r]{\strut{} 40}}%
      \put(946,1853){\makebox(0,0)[r]{\strut{} 60}}%
      \put(946,2236){\makebox(0,0)[r]{\strut{} 80}}%
      \put(946,2618){\makebox(0,0)[r]{\strut{} 100}}%
      \put(946,3001){\makebox(0,0)[r]{\strut{} 120}}%
      \put(946,3384){\makebox(0,0)[r]{\strut{} 140}}%
      \put(946,3767){\makebox(0,0)[r]{\strut{} 160}}%
      \put(1078,484){\makebox(0,0){\strut{} 0}}%
      \put(1935,484){\makebox(0,0){\strut{} 0.5}}%
      \put(2792,484){\makebox(0,0){\strut{} 1}}%
      \put(3649,484){\makebox(0,0){\strut{} 1.5}}%
      \put(4506,484){\makebox(0,0){\strut{} 2}}%
      \put(5363,484){\makebox(0,0){\strut{} 2.5}}%
      \put(176,2235){\rotatebox{-270}{\makebox(0,0){\strut{}$\tau/M$}}}%
      \put(3220,154){\makebox(0,0){\strut{}$x_{\rm g}/M$}}%
    }%
    \gplgaddtomacro\gplfronttext{%
    }%
    \gplbacktext
    \put(0,0){\includegraphics{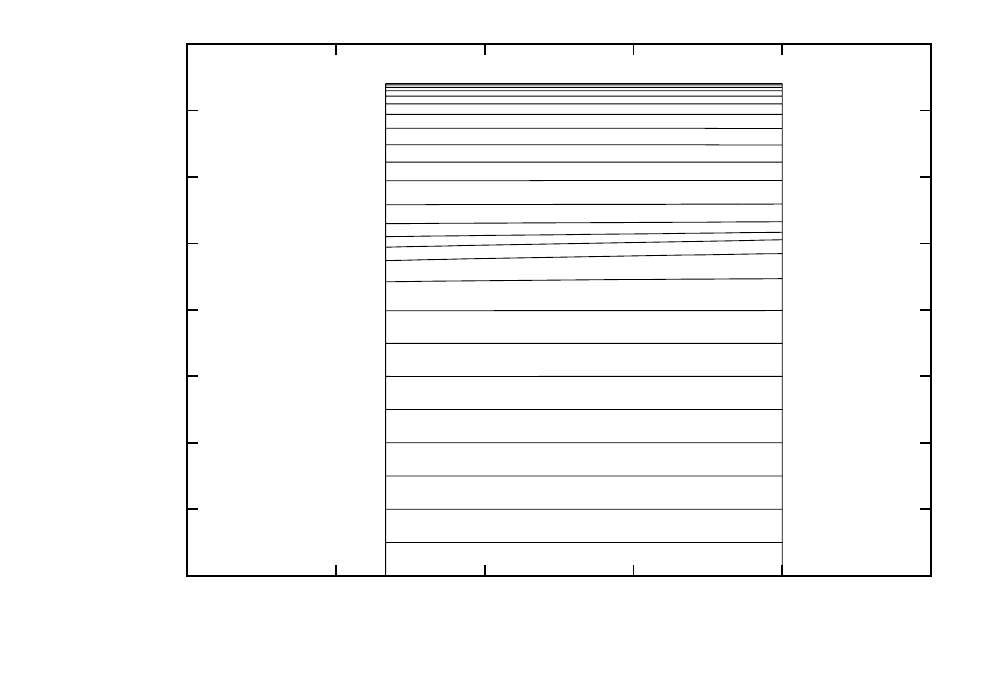}}%
    \gplfronttext
  \end{picture}%
\endgroup

\ece
\caption{Space-time region spanned by the edge between (2/3,2/3,2/3) and (2,2,2), represented in $(\tau,x_{\rm g})$ 
coordinates. The slicing adopted in this work only extends, in this region, up to $\tau \lesssim 150M$.
The horizontal lines are constant-$t$ lines, i.e.~sections of the spatial hypersurfaces used in the
simulation.\label{fig:numcoords}}
\efi

Figure~\ref{fig:pd} also includes a possible counterparts of the eight-black-hole lattice in the FLRW class:
one with the same initial edge length. This definition is made more
precise in section~\ref{sec:contrast}, where
we will constrast these results with the large-scale dynamics
of a perfectly homogeneous and isotropic universe.

Finally, let us notice here in passing that the interest of this toy model goes beyond 
the cosmological application.
In particular, it provides an interesting example of overlapping MOTSs
within the framework of a BSSN evolution.

\section{Comparison with the FLRW class}
\label{sec:contrast}

The comparison of the configuration with an FLRW model requires solving Ellis' ``fitting problem''~\cite{Ellis:1987zz}, i.e. determining the parameters characterizing the reference FLRW model which our configuration resembles most closely. There are infinitely many ways to do this; we will sketch below the procedure we will use in this paper based on the quantities measured in section~\ref{sec:evolution}.

Due to the symmetry group of our configuration, the reference FLRW model is a closed model ($k=1$), with spatial slices of spherical topology. The matter content is represented by dust.  
Since the primary variable describing any FLRW model is the scale factor, representing the scaling of lengths in time, the first step in the comparison is to identify some measure of length
in the lattice universe. We required this variable to be non-local, thereby capturing the large-scale, 
average behaviour of the universe rather than the local physics at a single, arbitrarily chosen point. The 
total volume of the configuration, which is the most obvious parameter for a closed universe, 
is obviously infinite and therefore of no use for our purpose. 

The problem of the size measurement is a bit simplified by the discrete symmetry of the model. In section~\ref{sec:evolution}, it seemed reasonable to choose the variable in a way which is consistent with the cell structure of the black hole lattice. The first obvious choice would be to use the geodesic distance $D_{\rm hor}(\tau)$ between the MOTSs of two neighbouring black holes. Recall that  a MOTS is a closed
two-surface whose null expansion vanishes in one direction~\cite{Andersson:2007fh}. We have observed that the behavior of this quantity for small times varies considerably from the behavior of the size parameter of a closed FLRW (see Figure~\ref{fig:pd}).
In particular, the derivative of $D_{\rm hor}(\tau)$ does not vanish at $\tau=0$, but rather approaches the limiting value $\dot D_{\rm hor}(0)=-2$, despite the configuration being momentarily
 static and symmetric with respect to time reflections. This striking phenomenon can be explained by  the fact that the 
 black hole MOTSs at $\tau=0$ are all bifurcation surfaces of the horizon. One can check that in our coordinates the corresponding MOTSs seem to
 approach each other at approximately the speed of light even at the moment of maximal expansion (see~\ref{app:horgeo}).
This makes $D_{\rm hor}(\tau)$ an unsuitable size parameter for the purpose of FLRW fitting.
  
Our second choice for the size variable was the geodesic length of the individual cell's edge, i.e. the geodesic distance between the two vertices of an individual cube, also used in \cite{Clifton:2012qh}. The edges lie relatively far from the punctures, and thus from the potentially problematic black hole region, at all times. Hence we can hope they are resolved quite well during the simulation. 

The mapping to the FLRW class is then carried out by fitting the size $a_{\rm{eff}}(t)$ of the reference sphere discussed in section~\ref{sec:S3punct} by demanding that the length of the edge match exactly the length of the corresponding edge of the cubical tiling of a round (FLRW) sphere. At the same time 
this fitting gives the effective Ricci curvature $\phantom{}^{(3)}R_\eff$ as the curvature of the corresponding FLRW sphere:
\begin{equation}
\phantom{}^{(3)}R_\eff = \frac{6}{a_\eff^2}.
\end{equation}

At first sight, it may seem quite strange that the effective Ricci curvature is not given by any kind of average, over a domain
in the constant time slice, of the local values of the curvature. Keep in mind however that in the FLRW class, due to the high symmetry, the value of the (constant) Ricci curvature is directly related to
infinitely many other parameters characterizing the geometry, for example the relation between the volume and the area of spheres, the angle deficits, the geodesic focusing at any point, and so on. Since the geometry of the discrete models is not homogeneous, these simple relations are lost. Nevertheless it
is not a priori clear \emph{which} of those parameters we should regard as a convenient inhomogeneous generalization of the spatial $\phantom{}^{(3)}R$ appearing in the Friedmann equation. The definition of $\phantom{}^{(3)}R_\eff$ we propose here is based on rescaling the value of the Ricci curvature of a unit 3-sphere by
the ratio of the size of the lattice edges. As we shall see, it provides, together with a complementary definition of the effective energy (see below), an excellent fit for the dynamics of the inhomogeneous lattice until up the time of $t \approx 80$, when we start losing resolution (mostly due to an outgoing shift vector that makes the system rapidly shrink in coordinate size). Independently from this limit,
after $\tau \sim 100M$ the eight MOTSs rapidly swallow up the whole spatial slices, making this model less and less appropriate to describe a universe filled with point-like masses.

In order to derive an effective energy density, let us recall that
the configuration is at rest at $t=0$, so we can assume that $\dot a_\eff(0)=0$ and use Friedmann's equation
to obtain $\rho_\eff$
\begin{eqnarray}
0 =  \frac{8\pi}{3}\rho_\eff - \frac{\phantom{}^{(3)}R_\eff}{6}.
\end{eqnarray}
It is instructive to compare the effective total mass, obtained as the product of $\rho_\eff $ with
the volume of the FLRW sphere, with the total matter content of our configuration:
\begin{eqnarray}
M_\eff = \rho_\eff \,2\pi^2a_\eff^3 = 378.78,\quad M_{8\rm{BH}} = 8 M_{\rm ADM} = 303.53
\end{eqnarray}
 Clearly the effective mass is around 25\% larger than the sum of ADM masses of the individual black holes. This is consistent
 with the expected nonlinear effects of gravitation, as the strong gravitational fields gravitate themselves.  It also agrees quite well with \cite{Clifton:2012qh}, where the fit with the corresponding FLRW model is based on imposing the equality of the total masses and the lengths of the edges are used for comparison instead.

\section{Conclusions}
\label{sec:conclusions}

We have discussed the construction of the initial data corresponding to periodic lattices of black holes using the Lichnerowicz-York construction. We have shown that just like in FLRW class there is a link between, on one hand, the curvature of the underlying constant-curvature metric (and thus also the topology of the spatial slice) and, on the other hand, the matter content and the Hubble parameter.  
In particular, it follows that it is not possible to find a flat, rectangular lattice of black holes without a momentary expansion or contraction.

We then recalled the construction of periodic lattices of black holes on an $S^3$ sphere with vanishing extrinsic curvature, originally introduced in~\cite{springerlink:10.1007/BF01889418}. We 
focused on the symmetric eight-black-hole configuration and showed that it can be conformally projected to an asymptotically-flat  seven-black-hole configuration with unequal mass parameters. We have discussed briefly its basic properties and then showed the results of the numerical evolution of this system. The main goal was the comparison with the time evolution of a dust-filled
closed FLRW model, evolved from the maximal expansion moment until the recollapse.

We proposed two methods for measuring the effective size of the configuration: the geodesic distance between the 
MOTSs of two neighbouring black holes and the geodesic length of the edge of a single cell. The first one turned out
to behave very differently from the size parameter of FLRW, as its derivative did not seem to vanish even at the initial slice. We explained this unusual behavior by noting that, at the maximum
expansion, each MOTS is a bifurcation surface where two distinct marginal tubes intersect. The edge length, on the other hand, calculated in normal coordinates and in proper time, seems to follow very closely the evolution of a closed FLRW if we fit its 
size (and consequently its curvature and mass) in an appropriate way. Despite our configuration being very far from homogeneity, the
effective size obeys the Friedmann equation to a remarkable degree up to times of $t\approx 80M$, which is approximately  30\% of the recollapse time of the FLRW. After that time our simulation is
simply unable to resolve the system. 
The only observable backreaction (or coarse-graining) effect in our simulation seems to
lie in the effective total mass of the system, which turns out to be 25\% larger that the sum of masses of the individual black holes. In other words, the eight-black-hole lattice
does mimic a closed FLRW dust model, but one whose total mass is substantially larger than that due to the black holes alone.

It is of course not clear to what extent these results will hold if we consider other types of models (flat or open, with a positive initial expansion) or if we drop the
assumption of existence of a large group of discrete symmetries. Note that at first sight the symmetry assumption may look like an innocent ansatz whose  only purpose is to simplify the geometry of the problem. Nevertheless
it is important to understand that it is in fact quite restrictive. In particular it prohibits many types of interactions
between the matter inhomogeneities, such as two-black-hole mergers or interactions via low-$\ell$ spherical harmonic modes.
Configurations which do not have that kind of symmetry may potentially exhibit many other effects of backreaction.

\section*{Acknowledgements}

We would like to thank Lars Andersson and Niall Ó Murchadha for useful discussions and comments. We also 
acknowledge the Erwin Schr\"odinger Institute in Vienna for its support and hospitality during the
2011 \emph{"Dynamics of General Relativity: Numerical and Analytical Approaches"} programme, where this
work was started. M.K.~acknowledges hospitality by the Max-Planck-Institut f\"ur Gravitationsphysik.
E.B.~is funded by a Marie Curie International Reintegration Grant PIRG05-GA-2009-249290.
Computations were carried out on the MPI-GP Damiana and Datura clusters.

\appendix
\section{Evolution system}
\label{app:bssn}
In order to solve Einstein's equation, we use the \texttt{McLachlan} code,
which implements a finite-difference discretization of the BSSN formulation
~\cite{Nakamura:1987zz, Shibata:1995we, Baumgarte:1998te}.
\bea
(\partial_t - \beta^l \partial_l) W &=& - \frac{1}{3} \alpha K + \frac{1}{3} \partial_i \beta^i \\
(\partial_t - \beta^l \partial_l) K &=& - D_i D^i \alpha + \alpha (\bar A_{ij} \bar A^{ij} + \frac{1}{3} K^2) \\
(\partial_t - \beta^l \partial_l) \bar \gamma_{ij} &=& -2 \alpha \bar A_{ij} + 2 \bar \gamma_{i(j}\partial_{k)} \beta^i - \frac{2}{3} \bar \gamma_{ij} \partial_k \beta^k\\ 
(\partial_t - \beta^l \partial_l) \bar A_{ij} &=& W^2 (-D_i D_j \alpha + a R_{ij})^{TF} \\ 
  && + \alpha (K \bar A_{ij} - 2 \bar A_{ik} \bar A^k_j) \\
  && + 2 \bar A_{k(i} \partial_{j)} \beta^k - \frac{2}{3} A_{ij} \partial_k \beta^k \\
(\partial_t - \beta^l \partial_l) \bar \Gamma^i &=& \bar \gamma^{jk} \beta^i \partial_j \beta_k + 
    \frac{1}{3} \bar \gamma^{ij} \partial_j \partial_k \beta_k - \bar \Gamma^j \partial_j \beta^i \\ 
  &&+ \frac{2}{3} \bar \Gamma^i \partial_j \beta^j 
      - 2 \bar A^{ij} \partial_j \alpha \\
  &&+ 2 \alpha (\bar \Gamma^i_{jk} \bar A^{jk} - 3 \bar A^{jk} \partial_k \ln W - \frac{2}{3} \bar \gamma^{ij} \partial_j K)
\eea
where $\gamma_{ij}=W^{-2} \bar \gamma_{ij}$ is the three-metric, $K_{ij}=\frac{K}{3} \gamma_{ij} + W^{-2} \bar A_{ij}$ is the
extrinsic curvature, and $W=\det\gamma^{-1/6}$ and $\bar \Gamma^i=-\partial_j \bar \gamma^{ij}$ 
are auxiliary variables.
The gauge variables are evolved according to:
\bea
(\partial_t - \beta^i \partial_i) \alpha &=& - 2 \alpha K \\
(\partial_t - \beta^i \partial_i) \beta^j &=& \frac{3}{4} B^j \\
(\partial_t - \beta^i \partial_i) B^j &=& - (\partial_t - \beta^i \partial_i) \bar \Gamma^j - \eta B^j
\eea
with $\eta=1$.

We use fourth-order finite-differencing.
An important difference with standard black-hole evolutions is that
we choose $\alpha=1$ everywhere
as the initial condition, because a precollapsed lapse $\alpha=\tilde \psi^{-2}$
(see (\ref{eq:confac}))
leads to a vast collapsed region at the center of the domain, which unnecessarily
slows down the proper-time evolution of the black holes. 

\section{Numerical error on the proper distance estimate}
\label{app:res}
Estimating the numerical error associated with a three-dimensional, complex
simulation is notoriously difficult. Additionally, our scheme for computing 
the proper distance does not only involve finite differencing and AMR operations,
but also a number of post-processing steps, in particular reslicing 1+1 spaces in terms of
gaussian observers (which involves using (\ref{eq:tgauss}) and (\ref{eq:xgauss}) to obtain the gaussian
coordinates, and then an interpolation in both time and space) and integrating
the line element to obtain a proper distance. 

In order to quantify the cumulative error of this procedure,
we evolve the same initial data at two additional resolutions, corresponding
to a spacing of $\Delta_0=2M$ and $\Delta=4M$ on the coarsest grid, and compare
the edge proper length to obtain an order of magnitude and 
a scaling for the numerical error on this quantity. The result is shown in Figure~\ref{fig:res}:
the error scales like the second power of the lattice spacing up until $t \sim 40M$, degrading 
afterwards. The order of magnitude is below $1M$ by $t=80M$ (or about $0.6\%$ of the proper
distance); this seems like a reasonable number to represent the numerics-related error bar on the
proper distance. 

\bfi
\bce
\begingroup
  \makeatletter
  \providecommand\color[2][]{%
    \GenericError{(gnuplot) \space\space\space\@spaces}{%
      Package color not loaded in conjunction with
      terminal option `colourtext'%
    }{See the gnuplot documentation for explanation.%
    }{Either use 'blacktext' in gnuplot or load the package
      color.sty in LaTeX.}%
    \renewcommand\color[2][]{}%
  }%
  \providecommand\includegraphics[2][]{%
    \GenericError{(gnuplot) \space\space\space\@spaces}{%
      Package graphicx or graphics not loaded%
    }{See the gnuplot documentation for explanation.%
    }{The gnuplot epslatex terminal needs graphicx.sty or graphics.sty.}%
    \renewcommand\includegraphics[2][]{}%
  }%
  \providecommand\rotatebox[2]{#2}%
  \@ifundefined{ifGPcolor}{%
    \newif\ifGPcolor
    \GPcolorfalse
  }{}%
  \@ifundefined{ifGPblacktext}{%
    \newif\ifGPblacktext
    \GPblacktexttrue
  }{}%
  \let\gplgaddtomacro\g@addto@macro
  \gdef\gplbacktext{}%
  \gdef\gplfronttext{}%
  \makeatother
  \ifGPblacktext
    \def\colorrgb#1{}%
    \def\colorgray#1{}%
  \else
    \ifGPcolor
      \def\colorrgb#1{\color[rgb]{#1}}%
      \def\colorgray#1{\color[gray]{#1}}%
      \expandafter\def\csname LTw\endcsname{\color{white}}%
      \expandafter\def\csname LTb\endcsname{\color{black}}%
      \expandafter\def\csname LTa\endcsname{\color{black}}%
      \expandafter\def\csname LT0\endcsname{\color[rgb]{1,0,0}}%
      \expandafter\def\csname LT1\endcsname{\color[rgb]{0,1,0}}%
      \expandafter\def\csname LT2\endcsname{\color[rgb]{0,0,1}}%
      \expandafter\def\csname LT3\endcsname{\color[rgb]{1,0,1}}%
      \expandafter\def\csname LT4\endcsname{\color[rgb]{0,1,1}}%
      \expandafter\def\csname LT5\endcsname{\color[rgb]{1,1,0}}%
      \expandafter\def\csname LT6\endcsname{\color[rgb]{0,0,0}}%
      \expandafter\def\csname LT7\endcsname{\color[rgb]{1,0.3,0}}%
      \expandafter\def\csname LT8\endcsname{\color[rgb]{0.5,0.5,0.5}}%
    \else
      \def\colorrgb#1{\color{black}}%
      \def\colorgray#1{\color[gray]{#1}}%
      \expandafter\def\csname LTw\endcsname{\color{white}}%
      \expandafter\def\csname LTb\endcsname{\color{black}}%
      \expandafter\def\csname LTa\endcsname{\color{black}}%
      \expandafter\def\csname LT0\endcsname{\color{black}}%
      \expandafter\def\csname LT1\endcsname{\color{black}}%
      \expandafter\def\csname LT2\endcsname{\color{black}}%
      \expandafter\def\csname LT3\endcsname{\color{black}}%
      \expandafter\def\csname LT4\endcsname{\color{black}}%
      \expandafter\def\csname LT5\endcsname{\color{black}}%
      \expandafter\def\csname LT6\endcsname{\color{black}}%
      \expandafter\def\csname LT7\endcsname{\color{black}}%
      \expandafter\def\csname LT8\endcsname{\color{black}}%
    \fi
  \fi
  \setlength{\unitlength}{0.0500bp}%
  \begin{picture}(7200.00,5040.00)%
    \gplgaddtomacro\gplbacktext{%
      \csname LTb\endcsname%
      \put(1861,704){\makebox(0,0)[r]{\strut{} 1e-05}}%
      \put(1861,1111){\makebox(0,0)[r]{\strut{} 0.0001}}%
      \put(1861,1518){\makebox(0,0)[r]{\strut{} 0.001}}%
      \put(1861,1925){\makebox(0,0)[r]{\strut{} 0.01}}%
      \put(1861,2332){\makebox(0,0)[r]{\strut{} 0.1}}%
      \put(1861,2740){\makebox(0,0)[r]{\strut{} 1}}%
      \put(1861,3147){\makebox(0,0)[r]{\strut{} 10}}%
      \put(1861,3554){\makebox(0,0)[r]{\strut{} 100}}%
      \put(1861,3961){\makebox(0,0)[r]{\strut{} 1000}}%
      \put(1861,4368){\makebox(0,0)[r]{\strut{} 10000}}%
      \put(1861,4775){\makebox(0,0)[r]{\strut{} 100000}}%
      \put(1993,484){\makebox(0,0){\strut{} 0}}%
      \put(2807,484){\makebox(0,0){\strut{} 20}}%
      \put(3621,484){\makebox(0,0){\strut{} 40}}%
      \put(4436,484){\makebox(0,0){\strut{} 60}}%
      \put(5250,484){\makebox(0,0){\strut{} 80}}%
      \put(6064,484){\makebox(0,0){\strut{} 100}}%
      \put(4028,154){\makebox(0,0){\strut{}$\tau/M$}}%
    }%
    \gplgaddtomacro\gplfronttext{%
      \csname LTb\endcsname%
      \put(5046,4626){\makebox(0,0)[r]{\strut{}$|D_{\rm c}(\tau)-D_{\rm m}(\tau)|$}}%
      \csname LTb\endcsname%
      \put(5046,4406){\makebox(0,0)[r]{\strut{}$F_1|D_{\rm m}(\tau)-D_{\rm f}(\tau)|$}}%
      \csname LTb\endcsname%
      \put(5046,4186){\makebox(0,0)[r]{\strut{}$F_2|D_{\rm m}(\tau)-D_{\rm f}(\tau)|$}}%
      \csname LTb\endcsname%
      \put(5046,3966){\makebox(0,0)[r]{\strut{}$F_3|D_{\rm m}(\tau)-D_{\rm f}(\tau)|$}}%
      \csname LTb\endcsname%
      \put(5046,3746){\makebox(0,0)[r]{\strut{}$F_4|D_{\rm m}(\tau)-D_{\rm f}(\tau)|$}}%
    }%
    \gplbacktext
    \put(0,0){\includegraphics{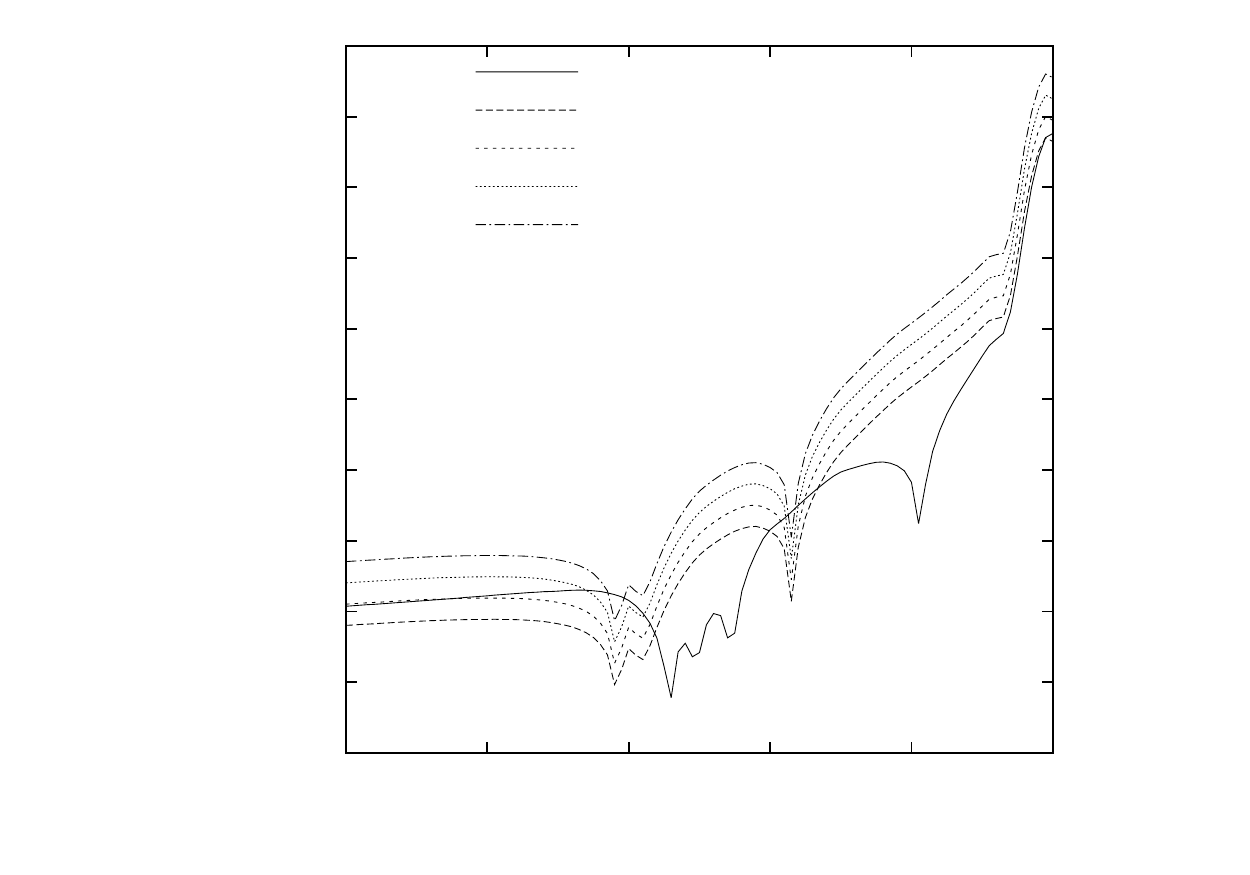}}%
    \gplfronttext
  \end{picture}%
\endgroup

\ece
\caption{Convergence study for the edge proper length. $D_{\rm c}(\tau)$, $D_{\rm m}(\tau)$, and 
$D_{\rm f}(\tau)$ are the edge proper lengths for the coarse, medium and fine resolution 
respectively. The factors $F_i$ represent $i$-th order scaling of the error with the lattice
spacing.\label{fig:res}}
\efi

\section{Geodesic distance between the MOTSs}
\label{app:horgeo}

In this appendix we will argue that the minimal surfaces around the punctures at $t=0$ are all bifurcation surfaces, i.e. they give rise to two MOTTs, one propagating towards the asymptotically flat end of the 
manifold and the other in the other direction.  Let $\Sigma_0$ be the $t=0$ hypersurface.
Consider the evolution of the initial data near $t = 0$ with gauge conditions $\alpha = 1$ and $\beta^i = 0$ at $\Sigma_0$, as we have chosen on our initial hypersurface for the numerical evolution.

\bfi
\bce
\includegraphics[width=7cm]{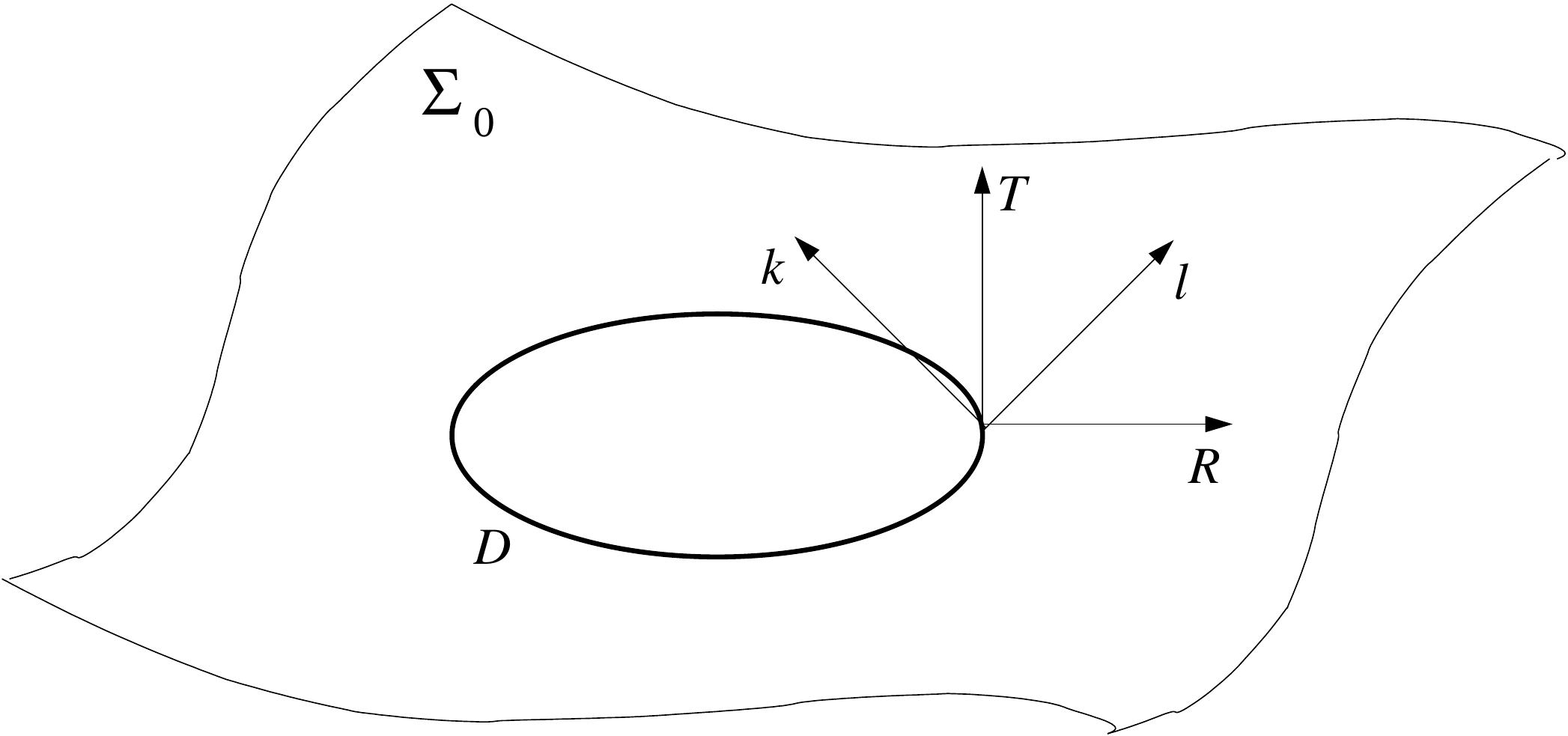}
\caption{Minimal surface $D$ as a subsurface of $\Sigma_0$ and its normals.\label{fig:normals}}
\ece
\efi

We define two null normals to $D$ given by $l = \frac{\sqrt{2}}{2}\left(T+R\right)$ and $k = \frac{\sqrt{2}}{2}\left(T-R\right)$, see Figure
\ref{fig:normals}. 
It is easy to see that the expansion of both null normals $\theta^l$  and $\theta^k$ vanishes. The reason for that is the fact that $D$ is minimal, therefore the expansion with respect to $R$ vanishes, and that the initial data is time-symmetric ($K_{ij} = 0$), so the expansion in $T$ direction
is zero as well.

The behavior of the MOTS is determined by the continuation equation for a MOTT, which gives
the condition under which a variation of a MOTS preserves the vanishing of one of the null expansions, see \cite{Korzynski:2006bx, Andersson:2007fh}. Let now $n = n_l\,l + n_k\,k$ be the variation vector field. Since \emph{both} null expansions vanish, we 
may continue the MOTS  by imposing either the condition $\theta^l = 0$ or $\theta^k = 0$. In the first case the continuation equation
reads
\begin{eqnarray}
\left(\Delta - 2\omega^A\,D_A - \left(\frac{{\cal R}}{2} + D_A \omega^A - \omega_A\,\omega^A\right)\right) n_k - n_l\,\left|\sigma_l\right|^2 = 0. \label{eq:MOTT}
\end{eqnarray}
$\Delta$ denotes here the Laplace operator on $D$, $\omega_A = \nabla_A k^a\,l_a$,   $\sigma_l$ is the shear of $l$ and ${\cal R}$ is
the scalar curvature of $D$.  

\bfi
\bce
\includegraphics[width=7cm]{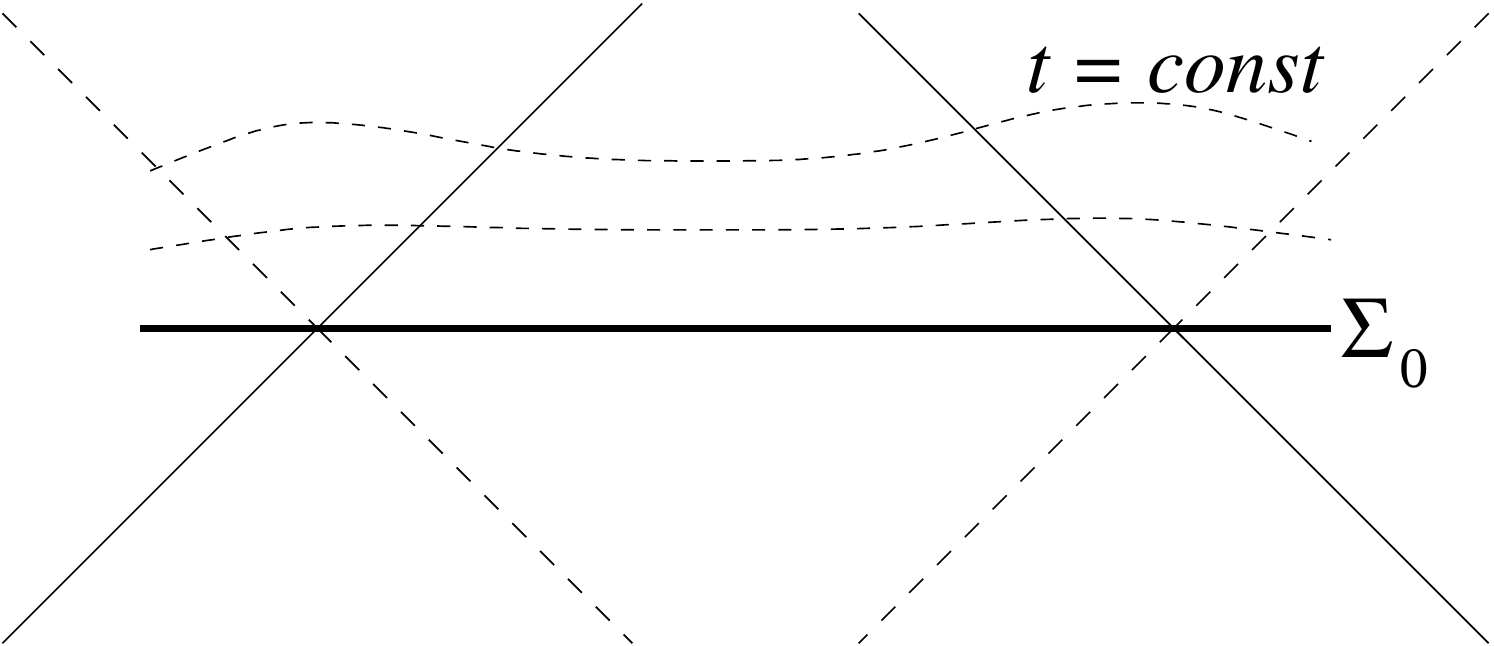}
\caption{Two neighbouring minimal surfaces at $t=0$ as bifurcation surfaces. The outward-moving MOTTs (solid line) seem to 
expand with approximately light speed.\label{fig:bifu}}
\ece
\efi

It turns out that both the surface $D$ itself and the physical 3-metric $\gamma$ in its vicinity are almost exactly spherically symmetric. The reason is that $D$ lies relatively close to the pucture $p$, at radius $r \approx 0.21$ in the $\RealNum^3$ variables, small comparing 
to the distance of  2 between the punctures. 
The round 3-sphere metric $\gamma^S$ is of course spherically symmetric around the puncture. The conformal factor $\psi$ is dominated in this region by the singularity at the origin and the sum of six terms coming from other punctures. The latter is not
exactly spherically symmetric, but the non-spherical contributions from those 6 terms  cancel 
out to a great degree due to their symmetric alignment. Therefore the relative variation of $\psi$ in angular variables at that 
place is only of the order of $10^{-5}$. All deviations from spherical symmetry are thus of very small order of magnitude, so in the first approximation we may neglect all terms in (\ref{eq:MOTT})
which break the spherical symmetry of $\gamma$, i.e. those containing the vector field $\omega$ and the tensor field $\sigma_l$~\footnote{Actually the shear term $|\sigma_l|^2$ in (\ref{eq:MOTT}), being quadratic, is of even lower order.}. The resulting equation now reads
\begin{eqnarray}
\left(\Delta - \frac{{\cal R}}{2} \right) n_k = 0.
\end{eqnarray}
The solution is straightforward: $n_k = 0$, $n_l$ being an arbitrary function. This obviously corresponds to the continuation of the MOTT along the null vector field $l$. The same argument shows that we may continue the other MOTT along $k$. Both MOTTs are thus in very good approximation isolated horizons.

It is now clear why the derivative of geodesic distance between MOTSs does not vanish despite the configuration being time reflection
symmetric. The MOTTs of two neighbouring black holes, when viewed  in either Gaussian or numerical coordinates, seem to expand almost exactly at the light speed (equal to 1 in simulation units). This by no means contradicts the time reflection symmetry of the inital data, as the MOTT expanding in the future direction is accompanied by another one which expands in the past direction, see Figure~\ref{fig:bifu}.
Figure~\ref{fig:speed}, showing the distance between MOTSs along with the expected initial slope of -2, provides support to the analysis above.

\bfi
\bce
\begingroup
  \makeatletter
  \providecommand\color[2][]{%
    \GenericError{(gnuplot) \space\space\space\@spaces}{%
      Package color not loaded in conjunction with
      terminal option `colourtext'%
    }{See the gnuplot documentation for explanation.%
    }{Either use 'blacktext' in gnuplot or load the package
      color.sty in LaTeX.}%
    \renewcommand\color[2][]{}%
  }%
  \providecommand\includegraphics[2][]{%
    \GenericError{(gnuplot) \space\space\space\@spaces}{%
      Package graphicx or graphics not loaded%
    }{See the gnuplot documentation for explanation.%
    }{The gnuplot epslatex terminal needs graphicx.sty or graphics.sty.}%
    \renewcommand\includegraphics[2][]{}%
  }%
  \providecommand\rotatebox[2]{#2}%
  \@ifundefined{ifGPcolor}{%
    \newif\ifGPcolor
    \GPcolorfalse
  }{}%
  \@ifundefined{ifGPblacktext}{%
    \newif\ifGPblacktext
    \GPblacktexttrue
  }{}%
  \let\gplgaddtomacro\g@addto@macro
  \gdef\gplbacktext{}%
  \gdef\gplfronttext{}%
  \makeatother
  \ifGPblacktext
    \def\colorrgb#1{}%
    \def\colorgray#1{}%
  \else
    \ifGPcolor
      \def\colorrgb#1{\color[rgb]{#1}}%
      \def\colorgray#1{\color[gray]{#1}}%
      \expandafter\def\csname LTw\endcsname{\color{white}}%
      \expandafter\def\csname LTb\endcsname{\color{black}}%
      \expandafter\def\csname LTa\endcsname{\color{black}}%
      \expandafter\def\csname LT0\endcsname{\color[rgb]{1,0,0}}%
      \expandafter\def\csname LT1\endcsname{\color[rgb]{0,1,0}}%
      \expandafter\def\csname LT2\endcsname{\color[rgb]{0,0,1}}%
      \expandafter\def\csname LT3\endcsname{\color[rgb]{1,0,1}}%
      \expandafter\def\csname LT4\endcsname{\color[rgb]{0,1,1}}%
      \expandafter\def\csname LT5\endcsname{\color[rgb]{1,1,0}}%
      \expandafter\def\csname LT6\endcsname{\color[rgb]{0,0,0}}%
      \expandafter\def\csname LT7\endcsname{\color[rgb]{1,0.3,0}}%
      \expandafter\def\csname LT8\endcsname{\color[rgb]{0.5,0.5,0.5}}%
    \else
      \def\colorrgb#1{\color{black}}%
      \def\colorgray#1{\color[gray]{#1}}%
      \expandafter\def\csname LTw\endcsname{\color{white}}%
      \expandafter\def\csname LTb\endcsname{\color{black}}%
      \expandafter\def\csname LTa\endcsname{\color{black}}%
      \expandafter\def\csname LT0\endcsname{\color{black}}%
      \expandafter\def\csname LT1\endcsname{\color{black}}%
      \expandafter\def\csname LT2\endcsname{\color{black}}%
      \expandafter\def\csname LT3\endcsname{\color{black}}%
      \expandafter\def\csname LT4\endcsname{\color{black}}%
      \expandafter\def\csname LT5\endcsname{\color{black}}%
      \expandafter\def\csname LT6\endcsname{\color{black}}%
      \expandafter\def\csname LT7\endcsname{\color{black}}%
      \expandafter\def\csname LT8\endcsname{\color{black}}%
    \fi
  \fi
  \setlength{\unitlength}{0.0500bp}%
  \begin{picture}(5760.00,4032.00)%
    \gplgaddtomacro\gplbacktext{%
      \csname LTb\endcsname%
      \put(946,704){\makebox(0,0)[r]{\strut{} 150}}%
      \put(946,1010){\makebox(0,0)[r]{\strut{} 160}}%
      \put(946,1317){\makebox(0,0)[r]{\strut{} 170}}%
      \put(946,1623){\makebox(0,0)[r]{\strut{} 180}}%
      \put(946,1929){\makebox(0,0)[r]{\strut{} 190}}%
      \put(946,2236){\makebox(0,0)[r]{\strut{} 200}}%
      \put(946,2542){\makebox(0,0)[r]{\strut{} 210}}%
      \put(946,2848){\makebox(0,0)[r]{\strut{} 220}}%
      \put(946,3154){\makebox(0,0)[r]{\strut{} 230}}%
      \put(946,3461){\makebox(0,0)[r]{\strut{} 240}}%
      \put(946,3767){\makebox(0,0)[r]{\strut{} 250}}%
      \put(1078,484){\makebox(0,0){\strut{} 0}}%
      \put(1614,484){\makebox(0,0){\strut{} 10}}%
      \put(2149,484){\makebox(0,0){\strut{} 20}}%
      \put(2685,484){\makebox(0,0){\strut{} 30}}%
      \put(3221,484){\makebox(0,0){\strut{} 40}}%
      \put(3756,484){\makebox(0,0){\strut{} 50}}%
      \put(4292,484){\makebox(0,0){\strut{} 60}}%
      \put(4827,484){\makebox(0,0){\strut{} 70}}%
      \put(5363,484){\makebox(0,0){\strut{} 80}}%
      \put(176,2235){\rotatebox{-270}{\makebox(0,0){\strut{}$D_{\rm hor}(\tau)/M$}}}%
      \put(3220,154){\makebox(0,0){\strut{}$\tau/M$}}%
    }%
    \gplgaddtomacro\gplfronttext{%
    }%
    \gplbacktext
    \put(0,0){\includegraphics{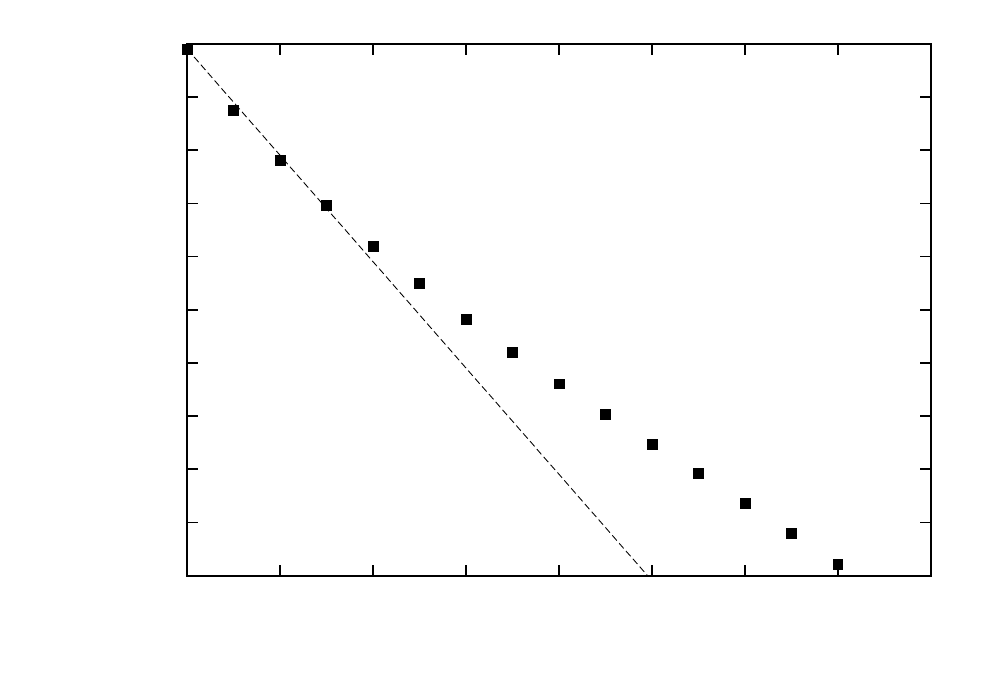}}%
    \gplfronttext
  \end{picture}%
\endgroup

\ece
\caption{The proper distance between near-neighbour MOTSs as a function of proper time. The straight
line has a slope of -2.\label{fig:speed}}
\efi

\section*{References}

\bibliographystyle{iopart-num}
\bibliography{eightBH}

\end{document}